\documentclass[prb,preprint,amsmath,amssymb]{revtex4}
\begin{document}
\author{Kevin Leung,$^{1*}$ Yue Qi,$^2$ Kevin R.~Zavadil,$^1$ Yoon Seok
Jung,$^{3,7}$ Anne~C. Dillon$^3$, Andrew S.~Cavanagh,$^4$ Se-Hee Lee,$^5$,
and Steven~M.  George$^6$}
\affiliation{$^1$Sandia National Laboratories, MS 1415 \& 0888,
Albuquerque, NM 87185\\
$^*${\tt kleung@sandia.gov}\\
$^2$General Motors R\& D Center. Warren, MI 48090, U.S.A.\\
$^3$National Renewable Energy Laboratory, Golden, CO 80401\\
$^4$Department of Physics, University of Colarado, Boulder, CO 80309\\
$^5$Department of Mechanical Engineering, University of Colarado, Boulder,
CO 80309\\
$^6$Department of Chemistry and Biochemistry, University of Colarado, Boulder,
$^7$Interdisciplinary School of Green Energy, Ulsan National
 Institute of Science and Technology (UNIST), Ulsan 689-798
 South Korea}
\date{\today}
\title{Using atomic layer deposition to hinder solvent decomposition
in lithium ion batteries: first principles modeling and experimental studies}

\input epsf
%\ssp
 
\begin{abstract}
 
Passivating lithium ion battery electrode surfaces to prevent electrolyte
decomposition is critical for battery operations.  Recent work on
conformal atomic layer deposition (ALD) coating of
anodes and cathodes has shown significant technological promise.  ALD
further provides well-characterized model platforms for understanding
electrolyte decomposition initiated by electron tunneling through
a passivating layer.  First principles calculations reveal two regimes
of electron transfer to adsorbed ethylene carbonate molecules (EC, a
main component of commercial electrolyte) depending on whether the electrode
is alumina-coated.  On bare Li metal electrode surfaces, EC accepts electrons
and decomposes within picoseconds.  In contrast, constrained density
functional theory calculations in an ultra-high vacuum setting show that,
with the oxide coating, $e^-$ tunneling to the adsorbed EC falls within the
non-adiabatic regime.  Here the molecular reorganization energy, computed
in the harmonic approximation, plays a key role in slowing down electron
transfer.  {\it Ab initio} molecular dynamics simulations conducted at
liquid EC-electrode interfaces are consistent with the view that reactions
and electron transfer occur right at the interface.
%although a widely used DFT functional
%is found to underestimate tunneling barriers.  
Microgravimetric measurements demonstrate that the ALD coating decreases
electrolyte decomposition and corroborate the theoretical predictions.
 
\end{abstract}
 
\maketitle
 
\section{Introduction}

Improving the fundamental scientific understanding of lithium ion
batteries\cite{book2,book,review} is critical for electric vehicles
and other energy storage technologies.  A key feature that enables the use
of negative electrodes (graphite, Li metal, Si, Sn) operating below the
reduction voltage of current commercial electrolytes is the formation of an
electronically passivating but Li$^+$-conducting solid electrolyte interphase
(SEI) film on electrode surfaces.\cite{book2,book,review,intro1,intro2}  
Battery performance, irreversible capacity ``loss,'' power fade,
durability, exfoliation of graphite, and safety are highly dependent on the
quality of the SEI.  Therefore understanding the nature, formation
composition, structure, and property of SEI is of great interest for
Li-ion batteries.  In this work, we apply computational and experimental
techniques to analyze the success of the conformal atomic layer deposition
(ALD) strategy for creating a passivating layer (``artificial SEI'') on
electrodes,\cite{dillon0,dillon1,dillon3,dillon4,dillon2} focusing on
graphitic carbon anodes.\cite{dillon} 

It is generally accepted that, upon the first charge of uncoated graphitic
anodes, the negative potential applied to induce Li$^+$ intercalation into
graphite decomposes ethylene carbonate (EC) molecules in the solvent, yielding
a self-limiting,
3-10~nm thick, passivating SEI layer containing Li$_2$CO$_3$, lithium
ethylene dicarbonate ((CH$_2$CO$_3$Li)$_2$),\cite{book,intro1,intro2} and salt
decomposition products.  Early modeling work on organic solvent breakdown has
focused on reactions inside bulk liquid regions, with an excess electron already
injected.\cite{bal01,bal,han,vollmer}  While providing extremely useful
predictions pertinent to that regime, such models necessarily ignore the
possibility of surface-assisted reactions and effects arising
from electron transfer from electrodes.  
A more rigorous if costly technique, {\it ab initio} molecular dynamics
(AIMD), has recently been applied to simulate chemical reactions at 
several explicit solid-liquid interfaces.\cite{marx,silica,sprik,sulpizi,car}
One of the authors' previous AIMD works follows chemical reactions in 
real time at the pristine graphitic anodes/liquid EC interface.\cite{ec,aimd}  
It is found that, at the initial stage of SEI formation, fast $e^-$ transfer
and kinetically-controlled EC electrochemical reactions occur to
form either CO or C$_2$H$_4$ gas,\cite{onuki,ota,aurbach_co,gachot}
mostly right at the oxidized edges of graphite sheets.\cite{ein,peled}

As electrolyte decomposition proceeds, $e^-$ transfer 
becomes impeded by the intervening and partially-formed 
SEI layer between the solvent and electrode, and the decomposed solvent
fragments can no longer anchor directly to the pristine electrode surface.
This important next stage should figure equally prominently in the overall
solvent breakdown mechanism and the structure of naturally-formed SEI.
The electron tunneling blockage by SEI layers is a {\it kinetic}
(not thermodynamic) phenomenon, akin to stoppage of electron leakage
through gate-oxide dielectric in semiconductor devices.\cite{high-k}  
Theoretical study there is hindered by the substantial thickness, 
possibly porous/gel-like nature, and heterogeneous composition of
natural SEI.\cite{book2,review,book,iddir}

Recently, it has been demonstrated that a sub-nanometer thick alumina layer
created by the conformal ALD technique on
graphite drastically diminishes solvent decomposition but permits lithium
ion transport.\cite{dillon}  This promising ALD strategy even enables the
cycling of low-melting-point propylene carbonate (PC), which otherwise
exfoliates and destroys uncoated graphitic anodes.  
The mechanism of this ALD electrode passivation has not been completely
understood.  While expected to block or slow down electron transfer
from the electrode to the solvent, it also appears to enhance the mechanical 
properties of the electrodes,\cite{dillon0} and likely hinders solvent
intercalation between graphite sheets, thus preventing exfoliation.  Apart
from the technological implications, the unprecedented control over coating
thickness and chemistry means that the ALD strategy also provides
robust platforms for basic science studies of interfacial solvent decomposition
reaction mechanisms, and for electron tunneling through the insulating
layer which is a pre-requisite for electrolyte breakdown.  

In particular, the extreme thinness of ALD coatings lends itself to the
present, predominantly first principles computational study of
electrode/solvent interfaces comprising up to 850~atoms.  Using
crystalline, hydroxylated LiAlO$_2$ layers as models of ALD
coatings, we apply DFT and related theoretical techniques to show that ALD 
oxide films yield varying energetic, kinetic, and electron-tunneling impedence
towards EC breakdown depending on the surface Li content and oxide thickness.
We also provide evidence that $e^-$ transfer occurs to
EC molecules immediately next to electrode surfaces.  Because these molecules
are deep within the electric double layer (EDL), screening of electric fields
by the EDL is less effective.  In this sense,
EC decomposition at battery anodes can differ fundamentally from 
classic electrochemical redox paradigms, where well-solvated transition
metal complexes are separated by several Angstroms from the electrodes
and ``outer-shell'' $e^-$ tunneling dominates.\cite{marcus}
The solvent decomposition processes on ALD coatings provide insights that may
be extrapolated to other passivation strategies, including natural SEI formed
from electrolyte breakdown.  

Two limiting regimes of electron transfer, and two corresponding
computational methods, are emphasized.
Rigorously, DFT deals with the electronic ground state, with
nuclear trajectories ``adiabatic'' to electronic configurations (i.e.,
ionic motions are slow compared with electron transfer).  EC breakdown on Li
(100) metal surface is in this adiabatic regime.  DFT should be
adequate for such processes, provided that the desired electronic
configuration is the ground state and the self-interaction
error of the approximate functional used is not critical to the
properties being investigated.\cite{wtyang,wtyang1,na,na1,deloc_ex}

\begin{figure}
\centerline{\hbox{ \epsfxsize=4.80in \epsfbox{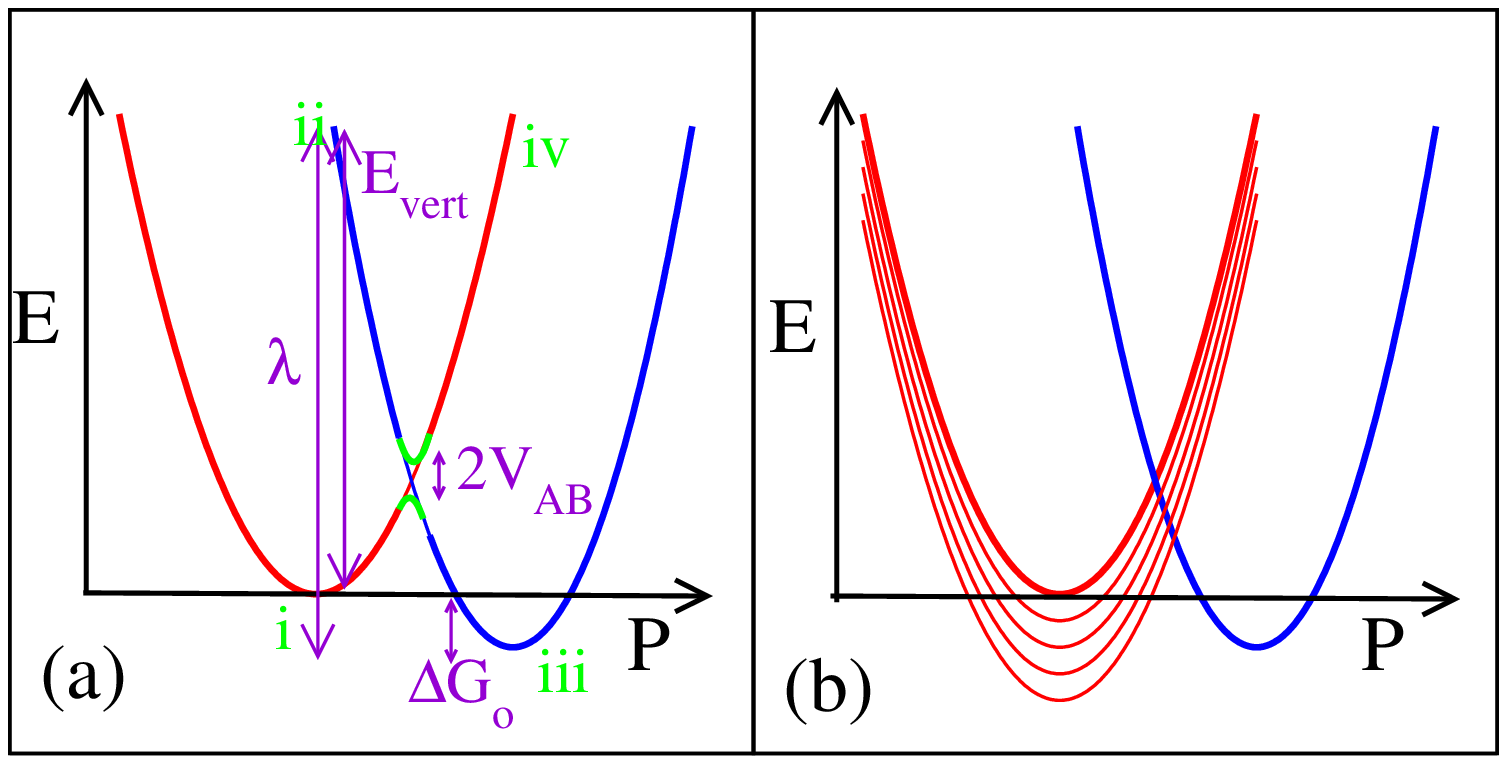} }}
\centerline{\hbox{ (b) \epsfxsize=2.00in \epsfbox{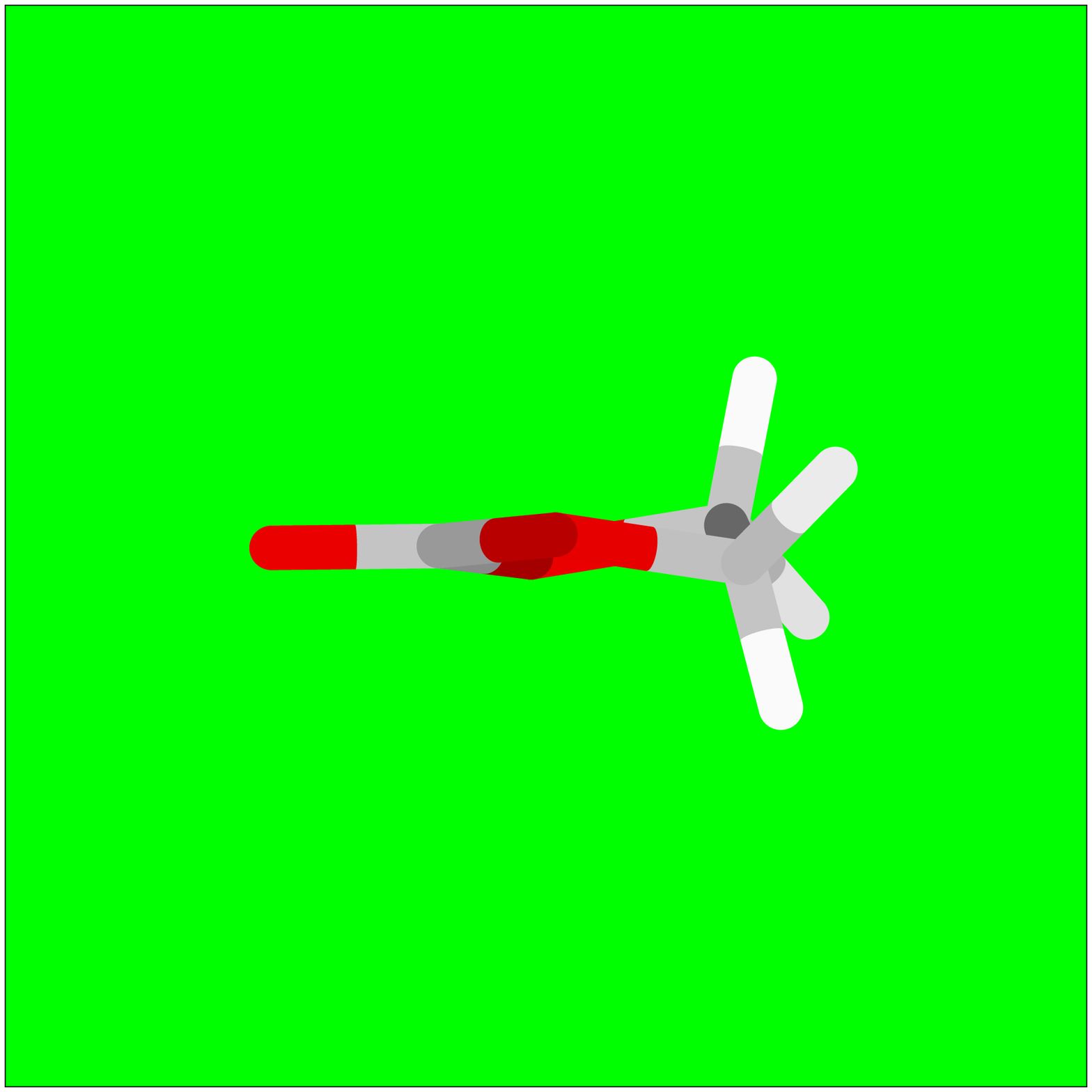}
                   (c) \epsfxsize=2.00in \epsfbox{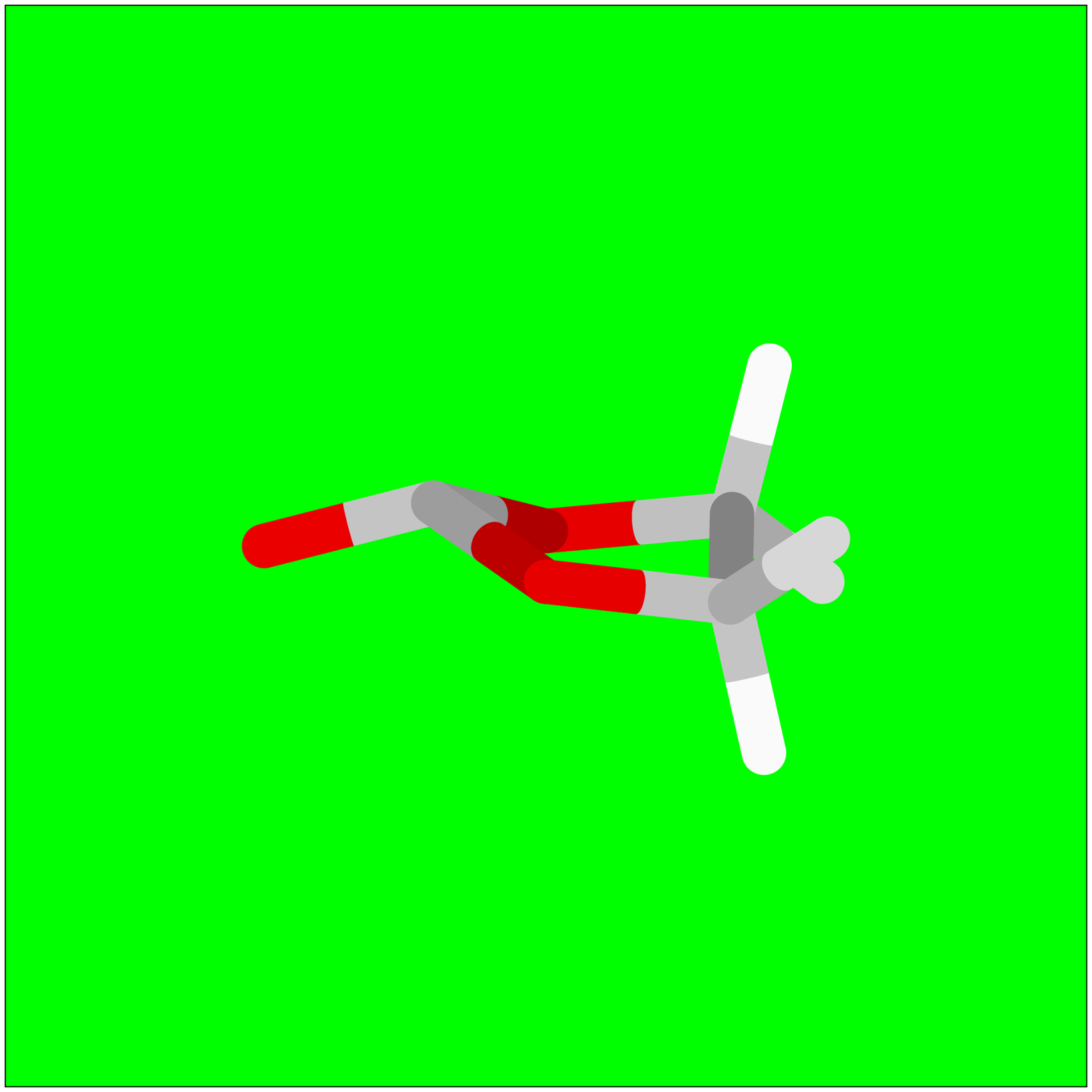} }}
\caption[]
{\label{fig1} \noindent
(a) Schematic of electron transfer between isolated orbitals.
The red and blue represent the diabatic potential energy
surfaces of $e^-$ donor and acceptor as a function of the
generalized polarization ($P$) degree of freedom.  The green
segments represent adiabatic processes with the non-crossing
surfaces split by $2 V_{\rm AB}$.  $\Delta G_o$ is the reaction
free energy and $\lambda$ is the reorganization energy.  The
green roman numbers denote (i) flat EC; (ii) flat EC$^-$; (iii)
bent EC$^-$; (iv) bent EC.
(b) Non-adiabatic $e^-$ transfer form a metallic electrode.
The thick upper red line represents the Fermi level, and is
the primary donor orbital within cDFT calculations.
$e^-$ can also transfer from the continuum of electrode donor
states below the Fermi level, depicted as thin red lines, to
the acceptor orbital, with however increased non-adiabatic
barriers (crossing points between blue and red curves).
(c) \& (d) Flat and bent EC molecules, respectively.  Red, grey,
and white refer to O, C, and H atoms.
}
\end{figure}

In the opposite, non-adiabatic
regime,\cite{tully,voor,voor1,newton,ulstrup} $e^-$ transfer
or tunneling is slow on the time scale of nuclear motion, and one must
keep track of two electronic surfaces.\cite{marcus}  The oxide-coated model
electrodes considered in this work pertains to this latter limit, where the
electron transfer rate between two discrete orbitals is given by\cite{dupuis}
\begin{equation}
k_{\rm et} = \frac{\sqrt{\pi}|V_{\rm AB}|^2 }{ \hbar
\sqrt { \lambda k_{\rm B}T} }
 \exp \bigg[ -\frac{(\Delta G_o + \lambda)^2}{4 \lambda k_{\rm B}T }\bigg].
\label{nonadiab}
\end{equation}
$\lambda$ is the reorganization (free) energy, $V_{\rm AB}$ is the coupling
matrix element connecting the two electronic surfaces, and $\Delta G_o$ is
the reaction free energy.  $\lambda$ indicates the energy cost
associated with molecular deformation needed to take on an extra electron
(EC $\rightarrow$ EC$^-$).  $V_{\rm AB}$ is the familiar prefactor
that depends on the overlap between two many-body
wavefunctions associated with the two electronic surfaces (Fig.~\ref{fig1}a).
Small $V_{\rm AB}$ correlates with non-adiabatic $e^-$ tunneling.

Neither $V_{\rm AB}$ nor $\lambda$ can be directly obtained using
standard DFT methods.  In this work, $\lambda$ is estimated using the
constrained DFT (cDFT) approach\cite{voor,voor2} and Marcus theory
harmonic construction (Fig.~\ref{fig1}a) under both ultra-high
vacuum (UHV) and liquid state\cite{blumberger2010,marzari} configurations.
cDFT is also applied to estimate $V_{\rm AB}$.\cite{voor1}.  While cDFT and
related methods have been applied to molecules on metal
surfaces,\cite{scheffler2008,scheffler2007,tully2009} calculating
$V_{\rm AB}$ between a metallic electrode and an $e^-$-accepting molecule,
or for that matter the total $e^-$ tunneling rate, has relied on simplified
models.\cite{dodonadze1968,schmickler1986,halley1988,sebastian1989,voth1995,voth1999,tanaka1999}  
When augmented using a Fermi Golden-rule expression (Fig.~\ref{fig1}b),
we argue that the our $V_{\rm AB}$ value yields a well-defined 
kinetic prefactor for electron transfer from a metallic electrode.  Our
prefactor prediction is a preliminary estimate, and fundamental studies to
extend cDFT to $e^-$ transfer from metallic electrodes are needed.  However,
this is sufficient for our goal of order-of-magnitude estimates of $e^-$
transfer rates.  When the insulating layer (ALD oxide or natural SEI,
or their combination) grows thicker, $V_{\rm AB}$ starts to decay with
oxide thickness, and its magnitude is examined via extrapolation in a way
analogous to the one-dimensional Wentzel-Kramers-Brillouin (WKB) formula.
An alternative to this cDFT formulation may be Greens function/time-dependent
DFT.\cite{rubio,sugino,roth} To our knowledge, TDDFT methods have not
been successfully applied to predict orbital-to-orbital $V_{\rm AB}$
values that involve metallic electrodes.

With these computational techniques, we show that the sub-nanometer
oxide coating,\cite{dillon,dillon1} generally not considered sufficiently
thick for complete electron blockage in, say, gate oxide dielectric
applications,\cite{high-k} causes $\lambda$ (much neglected in previous
battery studies) to play a significant role in ALD-assisted passivation.
Electron tunneling to EC, not bond-breaking
within the adsorbed molecule, is generally found to be the rate-determining
step for breakdown of EC adsorbed on the ALD-coated electrode.
%and addition of a second electron to EC with a broken bond is far easier than
%adding the first electron to an intact, adsorbed EC molecule.

%X-ray photoelectron spectroscopy (XPS) to determine film composition, and 
In terms of experiments, microgravimetric measurements that confirm
the presence of solvent decomposition products on the surface are presented
to corroborate aspects of our predictions.  

This paper is organized as follows.  Section~\ref{method} describes
the methods used.  Adiabatic electron-transfer induced EC reactions with Li
metal surfaces is discussed in Sec.~\ref{lithium}.  The long-range electron
transfer formalism is shown to be inapplicable here.  Section~\ref{oxide}
describes the non-adiabatic electron tunneling from oxide-coated electrodes
to EC molecules adsorbed on their surfaces, and addresses the subsequent
EC bond-breaking events.  Adiabatic DFT/PBE calculations are shown to
underestimate the electron tunneling barrier in this regime.
Section~\ref{expt} reports the experimental results, and
Sec.~\ref{conclude} briefly summarizes this work.  
 
\section{Methods}

\label{method}

\subsection{Model systems}
 
The casual reader is encouraged to skip forward to Sec.~\ref{lithium}
for the results.

The key model systems are $\sim$7.0~\AA\, (``thin'') and $\sim$10~\AA\,
(``thick'') layers of LiAlO$_2$ in $\beta$-NaAlO$_2$ structure with (100)
surface terminations, coated on narrow strips of Li$_x$C$_6$ electrodes
(Table~\ref{table1}; Fig.~\ref{fig2}; more details of this oxide phase
is provided in the supporting information (S.I.) document).  Undercoordinated
Al atoms on outer surfaces are terminated with OH groups, ensuring that
surface states are removed.  The oxide thickness is measured from Al 
to Al and excludes the surface hydroxyl groups or the C=O edge
atoms originally residing on Li$_x$C$_6$.  Crystalline LiAlO$_2$ is a
solid-state electrolyte candidate material.\cite{lialoh} LiAlO$_2$ is used
instead of Al$_2$O$_3$ to cover the possibility that the native Al$_2$O$_3$
layers deposited during ALD may have incorporated Li ions during the first
charging half cycle.  For example, some AlOH groups may be deprotonated at
low voltages, causing Li$^+$ to coordinate to the AlO$^-$ and become
part of the surface.  The LiAlO$_2$ mixed oxide
thus allows us to examine surface composition effects on EC breakdown.  The
stochiometry of the coating is such that their formal charges sum to zero.  
Another research group has found LiAlO$_2$ signature on the surface
of 5~nm thick ALD oxide films on Si anodes after power cycling using X-ray
photo-electron spectroscopy.\cite{xcx}  Further computational evidence
for Li-incorporation into Al$_2$O$_3$ films is presented
in the S.I.  If such Li$^+$ incorporation indeed occurs,
the ALD layer will expand beyond its original Al$_2$O$_3$ thickness.

The simulation cell, which provides a modest system size for AIMD simulations,
is chosen so that the oxide is fairly well-matched to the Li$_x$C$_6$
surface cell, with oxide compressive strains of 1.8~\% and 5.7\% in the
two lateral dimensions.  The crystalline models are idealized;
as in gate-oxide dielectric materials, insulating oxides should be amorphous
to minimize cracks.  The amount of Li present in the graphite region is
determined by tuning the Li chemical potential to 2.1~eV.  Upon geometry
optimization, Li ions initially residing at the C=O
edges become strongly coordinated to the bottom surface of the oxide coating.
Another model, with a single 10~\AA\, thick layer of LiAlO$_2$ hydroxylated
on both sides but no Li$_x$C$_6$ component, is used to examine post
$e^-$-transfer EC$^-$ bond-breaking.

To emphasize the influence of surface groups, we also include a model
with a $\sim$5.0~\AA\, thick layer of $\alpha$-Al$_2$O$_3$ coated on both
sides of the Li$_x$C$_6$ strip.  The oxide layers have (0001) terminations
with AlOH surface groups (Table~\ref{table1}).  
It has been predicted that $\gamma$-Al$_2$O$_3$ is more
stable than the $\alpha$ phase for film thickness below 36~\AA.\cite{ouyang}
However, this estimate was made without accounting for surface hydroxylation.
Since our thin Al$_2$O$_3$ film contains only two Al-O layers (not counting
the C=O edge groups), the oxygen positions are arguably consistent with
both $\alpha$-Al$_2$O$_3$ with closed
packed oxygen in ABAB stacking, and cubic $\gamma$-Al$_2$O$_3$ with ABCABC
stacking.  In $\alpha$-Al$_2$O$_3$, all Al are in octahedral sites while
Al occupy both octohedral and tetrahedral sites in $\gamma$-Al$_2$O$_3$.
Upon applying geometry optimization to the initial ``$\alpha$-Al$_2$O$_3$''
film, some Al ions are found to migrate to tetrahedral sites, especially
those coordinated to graphite-edge C=O groups.  Thus our
``$\alpha$-Al$_2$O$_3$'' film arguably exhibits both $\alpha$ and $\gamma$
character, consistent with experimental ALD coatings which are considered
amorphous without long-range order.

Finally, a thin slab of lithium metal truncated along (100) surfaces is
considered.  Even though Li metal itself cannot currently be used as
rechargeable anodes, EC breakdown products on Li are qualitatively
similar to those on LiC$_6$ surfaces.\cite{book2,bridel}  Under open circuit
conditions, Li metal should be at a well defined $\sim -3$~V versus
the standard hydrogen potential.\cite{solvation_li}  Furthermore,
EC decomposition on Li surface is free of the ambiguity associated
with solvent co-intercalation into graphite.\cite{besenhard}  Thus
Li metal provides an useful baseline with which to interpret predictions
for the oxide-coated surfaces.

\begin{table}\centering
\begin{tabular}{||l|l|l|l|l|l||} \hline
system/coating & method & Figure & stochiometry & cell size & N$_{\rm EC}$
							\\ \hline
thin LiAlO$_2$ & NEB & Fig.~\ref{fig4}c-d,~\ref{fig6} &
	Al$_{48}$Li$_{96}$O$_{148}$C$_{92}$H$_{24}$
                & 40.00$\times$12.47$\times$15.06 &  1\\
thick LiAlO$_2$ & NEB & NA &
        Al$_{72}$Li$_{118}$O$_{208}$C$_{92}$H$_{24}$
                & 46.0$\times$12.47$\times$15.06 &  1\\
thick LiAlO$_2$ & AIMD & Fig.~\ref{fig2}d &
        Al$_{72}$Li$_{118}$O$_{208}$C$_{92}$H$_{24}$
                & 48.5$\times$12.47$\times$15.06 &  36\\
thin LiAlO$_2$ & AIMD & Fig.~\ref{fig2}e &
	Al$_{48}$Li$_{96}$O$_{148}$C$_{92}$H$_{24}$
                & 43.00$\times$12.47$\times$15.06 &  36\\ \hline
only LiAlO$_2$ & NEB & NA &
	Al$_{36}$Li$_{36}$O$_{84}$H$_{24}$
                & 24.00$\times$12.47$\times$15.06 &  1\\ \hline
Al$_2$O$_3$ & AIMD & Fig.~\ref{fig2}f &
        Al$_{72}$O$_{204}$C$_{120}$H$_{72}$Li$_{51}$
                & 33.34$\times$14.97$\times$18.82 & 36 \\ \hline
Li (100) & AIMD & Fig.~\ref{fig2}b & Li$_{96}$ &
	30.35$\times$14.63$\times$14.63 & 32\\
Li (100) & NEB & Fig.~\ref{fig2}c & Li$_{96}$ &
	30.35$\times$14.63$\times$14.63 & 1\\
Li (100) & NEB & Fig.~\ref{fig3}b-d & Li$_{48}$ &
	24.00$\times$9.75$\times$9.75 & 1 \\ \hline
\end{tabular}
\caption[]
{\label{table1} \noindent
Details of systems used in AIMD and geometry optimization-plus-NEB
barrier calculations.  The spatial dimensions are in \AA.
``Stochiometry'' omits ethylene carbonate atoms in the liquid region.
}
\end{table}

\subsection{Adiabatic regime: DFT, AIMD simulations}

All calculations are performed using the Vienna
Atomic Simulation Package version 4.6 (VASP)\cite{vasp,vasp1}
and the PBE functional.\cite{pbe}  AIMD simulations apply 
$\Gamma$-point Brillouin zone sampling, a 400~eV planewave
energy cutoff, and a 10$^{-5}$~eV or 10$^{-6}$~eV convergence
criterion at each Born-Oppenheimer time step.  The trajectories
are kept at an average temperature of T=450~K using Nose thermostats,
except for the EC/Li metal simulation where T=350~K is enforced.
Tritium masses are substituted for protons to enable a time
step of 1~fs.  Under these conditions, the trajectories
exhibit drifts of less than 1~K/ps.   Due to the approximate
nature of DFT functionals and the simulation protocol (tritium
masses and thermostat used), the predicted reaction time scales
should be treated as relative, not absolute.  AIMD simulations 
reported do not account for spin-polarization.  Our previous
work has revealed no qualitative difference between
restricted singlet and spin-triplet DFT/AIMD simulations.\cite{ec}
Molecular configurations are pre-equilibrated using Monte Carlo
simulations and simple molecular force fields, as described in
an earlier work.\cite{ec}  Representative AIMD snapshots are depicted
in Fig.~\ref{fig2}.

The AIMD liquid/solid interfacial simulations are akin to dipping
electrodes fully pre-intercalated with Li into the organic solvent.
In principle, it may be possible to intercalate Li$^+$ in the electrolyte,
remove the anodes from solution, clean off possible decomposition
products in inert environments, and re-insert in solution to measure
the open circuit voltage.  Such experiments have not been performed
but can be attempted in the future.  In Sec.~\ref{work_func},
we further discuss the electrochemical potential of these electrode models.

T=0~K geometry optimizations and climbing image nudged elastic band
(NEB)\cite{neb} barrier calculations (e.g., Fig.~\ref{fig3}) are performed
with spin-polarization, a 10$^{-4}$~eV convergence criterion, and a linear
potential correction applied in the direction perpendicular to the surface
to remove dipole-image interactions.\cite{scheffler_sur} $\Gamma$-point
sampling is generally applied, except for calculations involving Li metal
slabs where 1$\times$2$\times$2 Brillouin zone sampling is used.  Even there,
$\Gamma$-point NEB calculations yield a C-O bond-breaking barrier
only 0.1~eV higher than the more dense Brillouin grid result.  It is
also found that the geometry and net charge of an adsorbed, intact EC$^-$ on
LiAlO$_2$ surface is unchanged whether $\Gamma$-point or 1$\times$2$\times$2
grids are used.  Comparing restricted singlet and spin-polarization results,
no difference is discernable in the EC on Li metal calculations, where the
bond-breaking barrier is small ($<0.1$~eV, Sec.~\ref{lithium}) and adiabatic
electron transfer from the electrode and the bond-breaking event occur
simultaneously.  These are the conditions under which spontaneously EC
decompositions are observed in picosecond AIMD simulations, justifying
the use of non-spin-polarized DFT there.  Higher bond-breaking barriers,
like those on the 10~\AA\, thick LiAlO$_2$ surface (see the S.I.), are
reduced when spin polarization is allowed.
A spot check shows that spin-unrestricted DFT calculations reduce the
C$_{\rm C}$-O cleavage barrier by 0.15~eV on this surface.  Even with this
reduction, the barrier is high enough to prevent observation of EC breakdown
in picosecond time scale, and therefore using non-spin-polarized DFT in
AIMD simulations does not affect the conclusion that no reactions occur
within the 7~ps trajectories in high barrier cases.  Further details on NEB
calculations are discussed in the S.I.

\subsection{Non-adiabatic regime: Constrained DFT}

A version of the constrained DFT (cDFT) method\cite{voor,voor1}
is implemented into VASP.  The constraining potential is chosen to be
\begin{eqnarray}
W({\bf r}) &=& V_o [ 1- \Pi_i f_i({\bf r}) ] \, , \label{v_o} \\
f_i({\bf r}) &=& [1+ \tanh ( \kappa (|{\bf r}-{\bf r}_i|-w_i)]/2  ,
\end{eqnarray}
Here $V_o$ is a constant to be self-consistently determined, $\kappa$ is
6\AA$^{-1}$, $i$ labels the atoms in the selected EC participating in
electron transfer, ${\bf r}_i$ is the atom position on that EC, and $w_i$
is an element-specific radius.  $w_i$ amounts to 1.65~\AA\, for C and O
and 1.25~\AA\, for H.  These values are similar to Lennard-Jones radii
in simple atomic force fields.  A more stringent wavefunction convergence
criterion of 10$^{-6}$~eV or smaller is enforced in self-consistent cDFT
calculations.  The $W({\bf r})$ functional form does not
double-count electron density on adjacent atoms and appears pertinent
when bond-breaking can occur.  
Normalized, atomic orbital-based charge projection operators used in
the literature\cite{voor,marzari,blumberger} may be less applicable for
electron transfer coupled to bond-breaking, but they can be tested
for the present application in the future.
%This is because $\int_{\bf r} W({\bf r}) \rho_e ({\bf r})$ is part of
%the cDFT total energy, where $\rho_e({\bf r})$ is the electron density,
%and this integral does not converge quadratically with wavefunction
%errors, making the higher precision necessary.

%$\lambda$ is computed using cDFT at the optimal 
%molecular geometry but with an electron enforced on the selected EC.  
%Alternatively, $\lambda$ can be obtained by removing an electron
%from the optimized electron-acceptor (EC$^-$) geometry.
%Within the assumptions of Marcus theory,\cite{marcus} the two
%should be equivalent. As will be discussed, in our UHV-like models,
%the former is the better choice.

The total electronic charge on the selected EC is determined by projecting
$W({\bf r})/V_o$ on to the DFT electron density.  With the $w_i$ values
mentioned above, unconstrained DFT predicts that a charge-neutral EC
molecule adsorbed on the thin LiAlO$_2$ surface (Fig.~\ref{fig4}c)
exhibits a slight $+0.20|e|$ ``net charge,''  while $-0.60$$\pm$0.1$|e|$
resides on the EC$^-$ (Fig.~\ref{fig4}d).  The adsorbed EC$^-$ exhibits a
similar $-0.67|e|$ charge on the LiAlO$_2$ oxide slab without any
conductive Li$_x$C$_6$ component (Table~\ref{table1}).  The non-integer values
arise because of residual charge densities at the edge of EC molecules
beyond the range of $W({\bf r})/V_o$.  (The net spin on EC$^-$ is 
about $0.9|e|$, and is more centered on EC than the net charge.)
Increasing $w_i$ is ruled out because of the close proximity of
adsorbed EC to the surface hydroxyl groups.  For example, using larger
$w_i$ has been found to lead to abstraction of protons from surface
hydroxyls.  The protons then bind to the negatively charged EC molecule.
Such reactions are not seen in unconstrained AIMD simulations and are deemed
unphysical.  We have therefore defined $+0.20|e|$ and $-0.60|e|$ to be
the net charges of flat, intact EC (Fig.~\ref{fig4}c) and EC$^-$
(Fig.~\ref{fig4}d) when using self-consistent cDFT calculations to
impose charges on the molecule.  
%Despite this, the charge on the EC$^-$ adsorbed on
%the thin LiAlO$_2$ surface should be unambiguous at low temperature,
%because decomposing Kohn-Sham orbitals on atomic centers shows that the
%majority spin EC orbital housing the extra electron (Fig.~\ref{fig4}a)
%is below the Fermi level.  
Increasing $|V_o|$ to increase the charge on adsorbed EC$^-$ to $-0.80|e|$
is found to yield only a 10\% change on the coupling matrix element
$V_{\rm AB}$, but can increase $\lambda$ by a fraction of an electron volt.
The more important parameter, the barrier in Eq.~\ref{nonadiab},
is only affected by $\sim \delta \lambda$/4 in the harmonic approximation
used in this work.  In the S.I., the predicted $\lambda$ for adsorbed
EC is shown to be comparable to that for EC in liquid EC, computed using
cluster calculations, localized orbitals, and a dielectric continuum
approximation.

Coupling matrix elements $V_{\rm AB}$ between the two different adiabatic
surfaces (Fig.~\ref{fig1}a) are computed using the cDFT formalism for
discrete orbital levels,\cite{voor1} which is implemented into VASP within the
projector-augmented wave formalism.\cite{vasp1} The same atomic configuration
must be used for both electronic surfaces, and this is chosen to be the
optimized atomic configuration where no excess electron resides on the flat,
adsorbed EC.  $V_{\rm AB}$ is generally assumed to be relatively independent of
atomic positions with the ``Franck-Condon'' approximation, although molecular
orientation dependence has been demonstrated.\cite{blumberger2010}
$V_{\rm AB}$ emerges from the 2$\times$2 Hamiltonian matrix $H$
connecting the donor ($|\Phi_A\rangle$, in our case from unconstrained DFT
calculations) and acceptor ($|\Phi_B\rangle$) single determinantal
wavefunctions.\cite{voor1,blumberger} $|\Phi_B\rangle$ features an excess
electron on one EC molecule and is generated using cDFT.  $H$ contains the
overlap matrix element $\langle \Phi_A | \Phi_B \rangle$ as well as 
$\langle \Phi_A | \sum_e W(r_e)|  \Phi_B \rangle$, where $e$ labels all
occupied electronic levels.\cite{voor1}  These calculations are fairly
costly and are performed at T=0~K in this work.

This cDFT-based $V_{\rm AB}$ formulation was originally devised for electron
transfer between ground state cDFT donar and acceptor electronic configurations,
with the implicit assumption that the relevant density-of-state is discrete.
In the limit of non-interacting electrons residing on a metal electrode, this
formalism reflects only the top curve on the left side of Fig.~\ref{fig1}b
and does not reduce to the well-known Fermi Golden Rule formula for tunneling
from a continuum of donor states.  Consider, in this limit, a band of
single-particle energy levels $E$ characterized by a density-of-state
$D(E)$ of orbitals $\phi(E)$, Fermi distribution function $f(E)$, Fermi
level $E_{\rm F}$, and an isolated acceptor orbital $\phi_a$ with energy
$E_a$.  The Golden Rule rate, associated with multiple level crossings
illustrated in Fig.~\ref{fig1}b, is
\begin{equation}
k_{\rm GR} \propto \bigg\langle \int dE |\langle \phi(E) | v(E) | 
 \phi_a \rangle |^2 D(E) f(E-E_{\rm F}) \delta (E-E_a) \bigg\rangle_{\bf R},
	\label{golden}
\end{equation}
where $v(E)$ is the single-particle coupling matrix element and $\langle O
\rangle_R$ denotes averaging over nuclear degrees of freedom $R$ on which
all quantities implicitly depend.  This formula allows many-electron acceptor
$|\Phi_{B'}\rangle$ states that involve $\phi_a$ but not the HOMO of
$|\Phi_A\rangle $, which represent electron-hole
excitations.\cite{sebastian1989}  In contrast, cDFT can only generate 
the electron-acceptor manifold $|\Phi_{B'}\rangle $ which is the ground
electronic states within the applied constraint.  

To incorporate the effect of Eq.~\ref{golden}, we make the common assumption
that $V_{\rm AB}$ is constant over the relevant range of density-of-
state.\cite{scheffler2008,tully2009,dodonadze1968,schmickler1986,halley1988,sebastian1989,voth1995,voth1999,tanaka1999}
Then an empirical Golden Rule-like expression can be proposed:
\begin{equation}
k_{\rm et}^{\rm GR} = \sum_{a'} f_{a'} \frac{\sqrt{\pi}|V_{\rm AB}|^2 }{ \hbar
 \sqrt {\lambda k_{\rm B}T} }
 \exp \bigg[ -\frac{(\Delta E_o + \Delta E_{a'}+ \lambda)^2}
		{4 \lambda k_{\rm B}T }\bigg] 
\label{et_final}
\end{equation}
Here $\Delta E_o$ is used in place of $\Delta G_o$ because we ignore 
entropy changes in T=0~K, UHV-setting calculations, $f_{a'}$ is the Fermi
and/or symmetry weight of Kohn-Sham orbital $a'$, and $\Delta E_{a'}$ is the
difference in energy between the Fermi energy and each Kohn Sham orbital
level $a'$, $e_{\rm F}$$-$$e_{a'}$.  $a'$ deep within the occupied
manifold does not contribute due to the $\Delta E_{a'}$ factor.
The self-consistent $\Gamma$-point electronic density is used to generate
a dense grid of occupied states $\phi_{a'}$ using a 1$\times$4$\times$4
Monkhorst-Pack Brillouin sampling.\cite{schultz}  

To converge to the infinite size limit for $e^-$ transfer to a single
EC molecule, the correct approach is not to increase
$k$-point sampling, but to increase all spatial dimensions of the
model electrode.  If the Li$_x$C$_6$ component of the electrode
is doubled in size in any one direction, the orbital donor
wavefunction $\phi_{a'}$ delocalized over the electrode is scaled
down by $\sim 1/\sqrt{2}$, and $|V_{\rm AB}|^2$ decreases 2-fold.
This underscores the fact that $V_{\rm AB}$ is not a measurable quantity
in finite-sized electrode models, but changes with the system size.  However,
the density-of-state $D(E_{a'})$ increases proportionately with system size,
and the sum over all orbital contributions (Eq.~\ref{et_final}) should be
well-defined in that infinite size limit.

\subsection{Experimental Details}

Carbon films deposited onto Cu were used as electrodes to explore the
passivating role of the ALD-derived alumina coatings with respect to
electrolyte reductive decomposition. Polished AT-cut quartz crystals
patterned with Cu electrodes (9~MHz, Inficion) were used as the base
current collector for conducting both voltammetry and gravimetry.
50 nm thick carbon films were deposited onto these crystals using a
pulsed laser deposition method.\cite{zavadil1,zavadil2} Conformal
alumina coatings were deposited onto
both carbon films and bare Cu electrodes at a substrate temperature of
180~$^o$C using alternate cycles of trimethylaluminum and water to
produce amorphous Al$_2$O$_3$ films of either 0.55 or 1.1~nm
thickness.\cite{dillon,ald1,ald2} Cycles of NO$_2$ and TMA pre-exposure
were used to ensure the nucleation and growth of a continuous alumina
film.\cite{ald3}
Electrochemical measurements were conducted under argon in a glove box
(Vacuum Atmospheres, $<$100~ppb~H$_2$O, $<$1~ppm O$_2$) in 1~M LiPF6
in a 1:1 volume mixture of ethylene carbonate and diethylcarbonate (Hoshimoto
and Kishida Chemical). A Solartron 1287 potentiostat coupled with a
Maxtek RQCM controller were used for simultaneous voltammetric and
gravimetric measurements. 
%X-ray photoelectron spectroscopy (XPS)
%was conducted on a Kratos AxisUltra instrument using a monochromatic
%Al(K) source. The crystals were transferred between the glove box and
%vacuum spectrometer using a transfer manipulator that ensured crystal
%exposure to only the glove box atmosphere (UHP Ar) and vacuum.

\section{Results: adiabatic AIMD/DFT predictions of EC/Li(100) reactions}
\label{lithium}

Adiabatic DFT/PBE calculations should be pertinent for the EC/Li(100)
interface, where EC and the metallic electrode are in close contact and
fast $e^-$ transfer is expected.
 
\subsection{Liquid EC on Li (100)}

Liquid EC has been previously predicted to decompose at the C=O edges
of LiC$_6$ electrodes within 7~ps at T=450~K in AIMD/PBE simulations.\cite{ec}
This timescale is used to qualitatively gauge the DFT/PBE predicted
reactivity of other surfaces towards liquid EC.

Figure~\ref{fig2}b shows that liquid EC decomposes readily on Li (100).
Within 15~ps, all 12 EC molecules adjacent to the Li metal, out of 32~EC
in the simulation cell, have accepted electrons and decomposed.  11 out
of these 12 exhibit two broken C$_{\rm C}$-O bonds to form CO +
OC$_2$H$_4$O$^{2-}$;\cite{onuki,ota,aurbach_co,gachot} only one EC decomposes
in the classic C$_2$H$_4$ + CO$_3^{2-}$ route hitherto widely accepted in
the literature, cleaving both C$_{\rm E}$-O bonds.\cite{book2,book} 
Here C$_{\rm C}$ and C$_{\rm E}$ are the carbonyl and ethylene carbon atoms,
respectively.  This finding is consistent with those in Ref.~\onlinecite{ec},
where both CO and CO$_3^{2-}$ products emerge at the interface between
liquid EC and pristine LiC$_6$ with oxidized edge groups.  This agreement
is significant because, by construction, the models used in
Ref.~\onlinecite{ec} exclude solvent co-intercalation cited in the
``3-dimensional'' SEI formation pathway.\cite{besenhard} Nevertheless,
fast EC decomposition and identical products are predicted on both pristine
graphite and Li metal surfaces, showing that such co-intercalation is
not necessary for SEI initiation.  

In the EC/Li trajectory, the temperature is thermostat at T=350~K, not
T=450~K, to avoid melting the solid Li.  Despite this, the heat generated by
the reactions and the incorporation of CO into the metal slab have caused
significant amorphization.  In Ref.~\onlinecite{bal}, the initial 200~fs of
this trajectory is examined in detail.  The bent EC geometry, with the carbonyl
C=O displacing out of the EC plane, is shown to be correlated with electron
transfer to EC, just like for the isolated EC$^-$ in solution
(Fig.~\ref{fig1}d).\cite{bal01,bal}  This bent geometry plays a
critical role in electron transfer and reorganization energy
calculations in LiAlO$_2$-coated surfaces (see below).
Our AIMD simulations have shown that OC$_2$H$_4$O$^{2-}$ can react with
2 CO$_2$ to form the main SEI organic product ethylene dicarbonate. 
Whether this product is deposited at the initial stage of SEI growth
depends on the availability of CO$_2$ and the solubility of the
decomposition fragments.\cite{harris}

\subsection{Isolated EC on Li (100)}
\label{uhv}
 
Remarkably, even a single EC molecule, in the absence of the liquid
environment which stabilizes its ionic breakdown products, still
decomposes on Li (100) surfaces to form CO + OC$_2$H$_4$O$^{2-}$ within
picoseconds.  (Fig.~\ref{fig2}c) This suggests that a simple T=0~K energy
profile calculation is relevant to EC decomposition.\cite{liquid}  
%DFT/PBE predicts that
%the lowest unoccupied molecular orbital (LUMO) of an EC held at a distance
%from Li metal lie at least 1 eV above the Li Fermi level.  
%As discussed in connection with EC reorganization energies above,
%molecular deformation is a necessary condition for electron donation to EC.

Figure~\ref{fig3}a compares the T=0~K energy profiles of the two modes
of excess electron-induced EC breakdown on Li metal.  They show that cleaving
the C$_{\rm C}$-O bond to form the precursor to carbon monoxide,
OC$_2$H$_4$OCO$^{2-}$, is thermodynamically less favorable than the
ethylene carbon-oxygen bonds to form CO$_3^{2-}$ and C$_2$H$_4$ by
a substantial 1.53~eV.  Cleaving the remaining C$_{\rm C}$-O bond
in the CO route only leads to another 0.16~eV stabilization.  The barriers
associated with both types of bond-breaking are vanishingly small.  Applying
the HSE06 truncated hybrid functional,\cite{hse06,hse06a} which exhibits
far less self-interaction errors\cite{wtyang,wtyang1,na} than PBE, 
increases the C$_{\rm C}$-O breaking barrier, but only to
0.16~eV (not shown, but consistent with the similar short time dynamics
predicted with the PBE and HSE06 functionals\cite{bal}).  This suggests
that adiabatic DFT/PBE barrier predictions are reasonably accurate for EC
in contact with Li metal.  The small barrier explains why both product
channels are available in picosecond time scales at explicit liquid
EC/electrode interfaces (Fig.~\ref{fig2}b, Ref.~\onlinecite{bal}).
We speculate that the kinetic prefactor favors the CO-route and makes it the
majority product in liquid-solid interface simulations (Fig.~\ref{fig2}b).

\subsection{Long-range e$^-$ transfer formalism is not applicable to EC/Li(100)}
\label{li_nonadiab}

For $e^-$ transfer to EC directly adsorbed on uncoated electrode surfaces,
the close contact should render the cDFT method for non-adiabatic
long-range electron transfer\cite{voor1,voor2} inapplicable.
If one insists on calculating $V_{\rm AB}$ using cDFT and
and the simulation cell described in Table~\ref{table1},
%the vertical excitation energy $(\lambda+\Delta E_o)$ (Fig.~\ref{fig1}a) is
%found to be 2.72~eV 
$V_{\rm AB}$ is found to be $0.23$~eV for a flat EC adsorbed on Li
(Fig.~\ref{fig3}b).  This large $V_{\rm AB}$ is consistent with
the significant, 56\% overlap between the acceptor and donor many-electron
wavefunctions, and should put the system in the adiabatic electron
transfer regime --- even with the caveat about the system size dependence
of $V_{\rm AB}$.\cite{landau}   (For comparison, a theoretical work on NO
molecules adsorbed on Ag(111), not using cDFT, has also yielded fraction-of-eV
$V_{\rm AB}$.\cite{tully2009})  We conclude that the adiabatic DFT/PBE
treatment should suffice in this case.

\section{Results: Non-adiabatic electron transfer to EC on oxide surfaces}
\label{oxide}

This section focuses on a UHV-like model consisting of an
isolated EC adsorbed on the lithium-intercalated graphitic carbon strip 
coated with LiAlO$_2$.  A 0.4~V/\AA\, electric
field is applied.  For this model, $e^-$ tunneling resides in the
non-adiabatic regime where cDFT calculations are pertinent.  
The relevance of this model to the liquid EC/electrode interfacial
environment will be clarified below.

\subsection{Two metastable EC charge states on 7~\AA\, thick oxide surface}

The Li$_x$C$_6$ model with a 7~\AA\, thick LiAlO$_2$ coating 
proves especially useful for examining the details of electron
transfer from the electrode to an adsorbed EC, which either precedes
or takes place simultaneously with EC$^-$ decomposition.   Two
(meta)-stable adsorbed EC configurations can be stabilized (Fig.~\ref{fig4}).  
One is a flat, charge-neutral EC coordinated to a surface site (an
AlOH group) via its carbonyl oxygen atom (Fig.~\ref{fig4}c).
Figure~\ref{fig4}a depicts the local electronic density-of-state
(DOS) for this system.  The Li$_x$C$_6$ region contains partially
occupied states near the Fermi level ($E_{\rm F}$).  The insulating
oxide spans a substantial band gap, although there are surface states
in the interface with Li$_x$C$_6$ that reduce the effective insulating
thickness.  The highest occupied molecular orbital (HOMO) of EC is below
$-2.5$~eV while the LUMO lies above $E_{\rm F}$.  This DOS is consistent
with a charge-neutral EC weakly interacting with the oxide surface.

The other configuration has an intact EC$^-$ which adopts a bent geometry
with the C=O bond protruding out of the EC plane (Fig.~\ref{fig4}d).
This is reminiscent of the first stage of liquid EC decomposition on
Li (100) surface, where the $e^-$-accepting EC adopts a similar bent
configuration.\cite{bal,note1} The excess charge on the EC is centered
around the carbonyl oxygen atom which is coordinated to two AlOH groups
and a Li surface atom.  The system exhibits a DOS
(Fig.~\ref{fig4}b) substantially different from Fig.~\ref{fig4}a.  The 
majority spin, highest occupied state of the EC molecule now lies below
the Fermi level.  The shift in the LUMO upon $e^-$ addition serves as
a caveat against using the LUMO of the neutral molecule as a figure-of-merit
in assessing electrochemical reduction tendencies.

The bent EC$^-$ is almost iso-energetic with the flat EC.  Its slight
exothermicity, $\Delta E_o=-0.02$~eV, does not depend on whether the electron
transfer is adiabatic or non-adiabatic.  It should not be affected by the
periodic images imposed by the simulation cell because the dipole correction is
applied.\cite{scheffler_sur} In fact, despite the transfer of an $e^-$ across
a 7- or 10-\AA\, thick oxide layer, the overall dipole moment of the
simulation cell changes by less than $1.0~|e|$\AA, apparently because the
electron density in the metallic Li$_x$C$_6$ strip can rearrange itself to
accommodate the electron transfer.  
The total charge in the simulation cell is conserved in these 
calculations and the large correction due to periodic boundary conditions
for isolated ions in solutions is not needed.\cite{seealso,sprik1}  
Note that $\Delta E_o$ is used in place of $\Delta G_o$ because the
calculation is performed at T=0~K.

\subsection{Non-adiabatic electron transfer on oxide surface}
\label{nonadiab1}

We apply the cDFT method
to calculate $\lambda$ and $V_{\rm AB}$ required to estimate the electron
transfer rate $k_{\rm et}$ (Eqs.~\ref{nonadiab} \&~\ref{et_final}).  We
stress that the flat EC absorbed on the oxide coatings is treated using
unconstrained DFT/PBE.  The highest-occupied orbitals of the 7~\AA\, and
10~\AA\, thick coatings reside in the Li$_x$C$_6$ region, and exhibit
integrated electron densities of less than 10$^{-4}$ and
5$\times$10$^{-8}$~$|e|$ on the EC molecule, respectively.  This shows that
the unconstrained DFT method already gives a reasonable description of
the neutral EC electronic configuration.

$\lambda$ is computed for the optimized, flat EC geometry adsorbed on the
thin LiAlO$_2$ coating (i.e. image~0 in Fig.~\ref{fig5}a).
cDFT imposes an extra electron on the EC molecule.  On the 7~\AA\, thick
coating, it yields a vertical
excitation energy $\Delta E_{\rm vert}$=$\lambda+\Delta E_o$=2.04~eV,
where $\Delta E_o$ is the aforementioned $-0.02$~eV offset between donor
and acceptor.  Alternatively, an electron can be removed from the
frozen bent EC$^-$ configuration (image~5), which
leads to $\lambda'+\Delta E_o$=1.80~eV. $\lambda$ and $\lambda'$
agree to within 14\%.  This is qualitatively consistent with the Marcus
theory postulate that the polarization degrees of freedom respond
harmonically (Eq.~\ref{nonadiab}), yielding a single reorganization
energy that governs electron transfer reactions.\cite{marcus}
With $\lambda$=2.06~eV for EC adsorbed on the thin LiAlO$_2$ surface, the
non-adiabatic barrier becomes 0.51~eV from a simple Marcus construction
(Eq.~\ref{nonadiab}).  This barrier is much higher than the $\sim$0.1~eV
adiabatic DFT/PBE activation energies for both the C$_{\rm C}$-O and
C$_{\rm E}$-O bond breaking pathways on this surface (Sec.~\ref{new_sec}),
and is therefore the rate-limiting step in EC breakdown on the surface of
the thin LiAlO$_2$ coating.

In the S.I., an EC with a dielectric approximation of the liquid EC solvent
medium is found to exhibit an average of $\lambda=1.76$~eV, similar to EC
adsorbed on the thin LiAlO$_2$ coating.  The co-solvent dimethyl carbonate (DMC)
exhibits only slightly smaller $\lambda$ values.  Therefore the substantial
$\lambda$, large compared to many organic molecules,\cite{bredas}
is intrinsic to out-of-plane bending of the C=O group as the
carbonyl carbon atom adopts a $sp^3$-like geometry to accommodate an $e^-$.
The S.I. further presents results on vertical excitation energy,
$\Delta E_{\rm vert}=\lambda+\Delta E_o$ (Fig.~\ref{fig1}a),
computed in several AIMD snapshots, to suggest that the Arrhenius term
in Eq.~\ref{nonadiab} favors $e^-$ transfer to EC molecules at the interface
over EC in the bulk liquid region.  In such AIMD simulations, we are limited
to the first choice of $\lambda$, i.e., instantaneously adding an electron
to EC, because EC$^-$ in liquid EC can have short lifetimes.\cite{ec}
Hence we will focus on this first choice throughout this work.

As this is a T=0~K calculation in a UHV-like setting, we have simply used the
$(\Delta E_o +\lambda)^2/(4 \lambda)$ expression in Eq.~\ref{nonadiab}
as the tunneling barrier,\cite{voor1,dupuis} and have not traced out the
two adiabatic curves as a function of the energy gap using liquid state
potential-of-mean-force simulations.\cite{wtyang2,blumberger1}
We have however checked that, when relaxing EC$^-$ 
frozen in the flat geometry (Fig.~\ref{fig4}c) with a constrained charge,
it reverts to the stable bent EC$^-$
(Fig.~\ref{fig4}d) configuration, showing that the cDFT approach puts
the system on the correct electron-acceptor potential surface.
In the future, we plan to perform direct cDFT calculation of the barrier
height at T=0~K by simultaneously optimizing the same atomic configuration
on both energy surfaces.

The cDFT coupling matrix element is estimated to be $V_{\rm AB}$=0.022~eV
at the flat EC geometry.  
%With this value, the transmission coefficient
%(Eq.~\ref{transmission}) $\kappa$=0.14.  This value, much less
%than unity, marks the onset of non-adiabaticity.  It is also 
%This value is arguably consistent with 0.01-0.2~eV computed for polaron
%hopping between Ti ion sites located $\sim 3$\AA\, apart in TiO$_2$
%crystals.\cite{dupuis,note99}   
Fig.~\ref{fig5}c depicts the highest-occupied DFT and cDFT orbitals,
integrated over the lateral dimensions, for the systems with flat EC
and flat EC$^-$ respectively.  The overlap between them,
$\langle \phi_{\rm EC}^{\rm HOMO}|\phi_{\rm EC-}^{\rm HOMO}\rangle$, is
0.0125, or within 2\% of that between the respective determinantal
wavefunctions $\langle \Phi_{\rm A}|\Phi_{\rm B}\rangle$ (Sec.~\ref{method})
which includes many-electron contributions.  Therefore the relaxation of
other electrons (``polarization effect'') does not strongly influence
the overlap integral when using $\Gamma$-point sampling.
%dG=0.00:   3755.81519  3.43168824E+12  1.09445116E-09
%dG=0.35:   1.86680705  3.43168824E+12  5.43990863E-13

This estimate of $V_{\rm AB}$ does not reflect the classic Fermi Golden
rule phenomenology (Sec.~\ref{method}).  Applying Eq.~\ref{et_final} to
approximately account for the finite density-of-state on the electrode, we
obtain a 1.63$\times$10$^4$/s electron transfer rate.  Simply using the cDFT
definition of $V_{\rm AB}$ in Eq.~\ref{nonadiab}, which represents a single
point integration quadrature, merely underestimates this rate by a factor of
1.68.  Using DFT/PBE rather than more accurate but costly hybrid functionals
has been known to overestimate $V_{\rm AB}$ by almost a factor of
10.\cite{blumberger2010} In the present case, the DFT/PBE underestimation of
the band gap of the insulating oxide layer may lead to some overestimation of
the electron tunneling rate.  Despite the approximations and
assumptions involved, this is to our knowledge the first DFT-based estimate
of the tunneling rates from an electrode, through an oxide layer, to an
adsorbed EC molecule.  The value may potentially be compared with UHV
measurements.  After electron transfer, EC$^-$ decomposes, and the negatively
charged EC fragments will most likely complex with Li$^+$ from the
electrolyte and be incorporated into the SEI layer on top of the ALD film.

\begin{table}\centering
\begin{tabular}{||l|l|l|l|||} \hline
\multicolumn{2}{|c|}{coated electrode} & 
\multicolumn{2}{|c|}{uncoated electrode} \\ \hline
system & work func. & system & work func. \\ \hline
thin LiAlO$_2$ (OH) & 2.47 & Li(100) & 3.05 \\
thick LiAlO$_2$ (OH) & 2.90  & graphite edge  & 4.57 \\
thin LiAlO$_2$ (OLi) & 2.25 & LiAlO$_2$ & 5.42 \\
Al2O3 (OH) & 4.10 & Al$_2$O$_3$ (0001) & 6.22 \\
\hline
\end{tabular}
\caption[]
{\label{table2} \noindent
Work function of model systems used in this work 
computed using the PBE functional, in eV.  The left column
describes the oxide coatings on Li$_x$C$_6$; the right column
refers to work functions of pure crystals.  The Al$_2$O$_3$ (0001)
model is Al-terminated.  The graphite slab has dangling bonds.
}
\end{table}

On the thicker LiAlO$_2$ coating, $\Delta E_o$=0.77~eV.
$\lambda$=1.98~eV is predicted in the flat EC geometry.  Removing
an $e^-$ from the bent geometry yields $\lambda'=1.69$~eV.
The asymmetry is 15\%.  We again adopt the first choice of $\lambda$.
$V_{\rm AB}$ is estimated at 0.0128~eV, about half that of the 7~\AA\,
thick LiAlO$_2$ coating.\cite{note77} As discussed in Sec.~\ref{oxide},
the thinner coating exhibits substantial surface relaxation which is absent
in the 10~\AA\, layer, making a purely thickness-based comparison of
$V_{\rm AB}$ difficult.  Fig.~\ref{fig5}d depicts the donor and acceptor
Kohn-Sham and cDFT orbitals.  The overlap between them,
$\langle \phi_{\rm EC}|\phi_{\rm EC-}\rangle$,
is about 0.004, a factor of 3 less than that across the 7~\AA\, thick coating.
Including the contributions of Eq.~\ref{et_final}, $k_{\rm et}$ becomes
extremely small (2.8$\times$10$^{-5}$/s) due to the larger $\Delta E_o$.
%If we impose $\Delta G_o=0$, and including the Golden rule augmentation
%(Eq.~\ref{et_final}), $k_{\rm et}$=1.4$\times$10$^3$/s.
%dG=0:   562.265416  8.47696165E+11  6.63286493E-10

The overall $k_{\rm et}$ is clearly very sensitive to $\Delta E_o$ or
$\Delta G_o$.  In UHV settings, $\Delta E_o$ depends on both the electric
field and surface heterogeneity at atomic lengscales (see below).  At
electrode/liquid electrolyte interfaces, $k_{\rm et}$ is a function of the
applied voltage as well as the local EC reduction potential via
$\Delta G_o$ (Eq.~\ref{nonadiab}), which may be a function of the distance
from the electrode.  Direct measurement of the reduction potential of
an intact EC is unavailable because EC decomposition occurs faster 
than cyclic voltammetry time scales.  If one adopts a theoretical
$\Delta G_o$=$-0.15$~eV for $e^-$ transfer to intact EC molecules
in EC liquid in at Li(s)/Li$^+$ voltages (S.I.), the predicted initial
electron transfer rates $k_{\rm et}$ through the 7-\AA\, and 10-\AA\, thick
oxide coatings ($\sim 1.7$$\times$$10^5$/s and $8.3$$\times$$10^4$/s),
will permit electrolyte breakdown, even if we assume that these rates are
overestimated by 100~times due to the use of the PBE functional discussed
above.  Indeed, our gravimetric measurements reveal electrolyte decomposition
on the coated electrodes --- consistent with ready availability of electrons
--- albeit in much less quantity than on uncoated electrodes (Sec.~\ref{expt}).
The electrolyte decomposition product then yields an additional insulating
layer that prevents further electron tunneling. 

Our main point in this section is not to predict exact $k_{\rm et}$
values, but to highlight the previously neglected role of the EC reorganization
energy ($\lambda$) on electrode coated with an insulating layer.  An immediate
implication is that different solvent molecules/salt components may exhibit
different $\lambda$ and $e^-$-transfer rates.

\subsection{DFT/PBE treatment of electron transfer on oxide surface is
inadequate}
\label{new_sec}

We next demonstrate that adiabatic DFT/PBE calculations are inadequate
when dealing with $e^-$ tunneling through insulating oxide layers.

The electron transfer barrier strongly depends on whether the $e^-$ transfer
is adiabatic or not, and on the accuracy of the DFT method used.
Figure~\ref{fig5}a depicts a climbing-image NEB calculation with
4~images along the reaction coordinate linking the flat EC and the bent EC$^-$ 
to examine the DFT/PBE adiabatic energy landscape in the 0.4~V/\AA\, electric
field.  DFT/PBE predicts a 0.09~eV barrier associated with electron transfer
through the thin LiAlO$_2$ layer.  

This small 0.09~eV value gives the strongest indication that
DFT/PBE grossly underestimates the $e^-$ transfer barrier.
In classical electron transfer paradigm (Fig.~\ref{fig1}a), the
parabolic intersection which yields the non-adiabatic barrier in the
exponential term in Eq.~\ref{nonadiab} is expected to differ from an 
adiabatic prediction of barrier by $V_{\rm AB}$.  Instead, the former is
0.51~eV and the latter is 0.09~eV (Fig~\ref{fig5}a); their difference far
exceeds $V_{\rm AB}$=0.022~eV before considering system size
dependence.  The discrepancy is most likely due to the
self-interaction error in the DFT/PBE functional,\cite{wtyang,na}
a point already alluded to in Ref.~\onlinecite{voor1}.  The widely used PBE
functional, along with others, do not sufficiently penalize configurations
where an electron occupies both the electrode and the EC molecule.  Indeed,
in image~2 of Fig.~\ref{fig5}a, a fractional $-0.2$~$|e|$ charge 
develops on the EC, which should be considered unphysical for a molecule
separated from the electrode by at least 7~\AA.  Hybrid DFT functionals
exhibit less self-interaction errors than DFT/PBE, but are currently
too costly for computing barriers in interfacial systems of
this size.\cite{na1}

The 10~\AA\,-thick oxide-coated electrode exhibits a monotonic
DFT/PBE energy profile for electron transfer.  There is no DFT/PBE
adiabatic barrier between the flat EC and bent EC$^-$ beyond the minimal
0.77~eV mandated by the endothermicity (Fig.~\ref{fig5}b), suggesting that
the electron tunneling barrier is again severely underestimated.  Using 
the conjugate gradient geometry minimizer in VASP, the bent EC$^-$
geometry on this surface is in fact on the verge of instability, about to lose
electron density to the electrode and relax to the flat EC$^0$ geometry.
Therefore the depicted energy profile actually reflects an optimized
geometry subject to a charge constrained via cDFT with a small $V_o=-0.2$~eV.

\subsection{Work function and electrochemical potential}
\label{work_func}

The electron tunneling rate at electrolyte-electrode interfaces depends on
the electrochemical potential ($\Phi$) of the electrode.  In the coated
graphite model systems, $\Phi$ is not precisely known.  Directly calculating
$\Phi$ involves averaging the electrostatic potential difference between
the conductive (inner) region of the electrodes and a distant point in
the bulk liquid beyond the thickness of the electric double layer,\cite{halley}
and involves consideration of image charge and surface potential
effects.\cite{sprik1,lynden1,pratt,sur_pot}  These are beyond the time and
length scales of current AIMD simulations.  Fortunately, the EC/Li(100)
interface mimics immersing freshly prepared Li metal into liquid EC,
and reflects an unambigous open-circuit voltage below the threshold at
which EC becomes electrochemically decomposed (+0.8~V vs.~Li$^+$/Li(s)).
This is a major reason Li is considered in this work.  

If we consider the energy of an $e^-$ in the bulk electrolyte to be a
constant, independent of electrode surfaces, the energy for ejecting an
electron from different electrodes into the
bulk electrolyte will only be shifted by the work function\cite{lang}
(where an $e^-$ goes into vacuum).  Thus, we have 
computed the work functions of coated and uncoated electrode surface
and some crystal planes of ALD coating materials (Table~\ref{table2}).
The -OH and -OLi terminated LiAlO$_2$ coating work functions
are within 0.5~eV of the Li metal value, indicating that similar 
energies are required to remove an electron from these surfaces.
The Al$_2$O$_3$ coated surface has a much higher work function
(Table~\ref{table2}), consistent with our observation that Al$_2$O$_3$
is a more insulating material than LiAlO$_2$ (see below).

Even though our DFT calculations show that placing these oxides in contact
with Li metal surfaces leads to immediate Li metal oxidation, we use Li(100)
as a reference because its voltage is similar to that of LiC$_6$.
Aligning the work functions of Li(100) and the oxide materials
(Table~\ref{table2}), it is clear that the valence and conduction bands
of the ALD phase lies below and above the Fermi energy ($E_{\rm F}$)
of Li metal, respectively.  Electron tunneling from the Li $E_{\rm F}$
to the conduction bands of Al-terminated Al$_2$O$_3$ (0001) and LiAlO$_2$
(100) exhibit 1.43~eV and 1.13~eV offsets (barriers, $\Delta E$), respectively.

According to the 1D WKB formula, the tunneling prefactor is 
\begin{equation}
k_{\rm et} \propto \exp (-2 \sqrt{2 m_e \Delta E} R/\hbar),
\end{equation} 
where $m_e$ is the electron mass.  If we take a tunneling transmission
probability of e$^{-40}$ as the limit of vanishing electron tunneling,
3.7nm thick LiAlO$_2$ and 3.2nm thick Al$_2$O$_3$ are required to stop
total SEI growth using DFT/PBE predicted $\Delta E$.  The work
function is only one contribution to $\Phi$ and does not contain
solvent orientation and electric double layer effects\cite{note67}
(which should be less important for our inner-shell redox reduction of
solvent compared to the classical paradigm of electron transfer to
well-solvated outer-shell ions).  Nevertheless, it gives a simple
guidance for comparing different insulating ALD coating materials. 
As $e^-$ transfer slows down and becomes rate-limiting, the
composition of SEI films formed from electrolyte decomposition will likely
change.  This is because solvent molecules (other than EC), the
counter ions (PF$_6^-$) in the salt, and other partially decomposed
products may exhibit smaller electron transfer barriers (reorganization
energies) and start dominating the product channel.

\subsection{EC bond-breaking on ALD coating after $e^-$ transfer}
\label{break}

\begin{figure}
\centerline{\hbox{ (a) \epsfxsize=2.00in \epsfbox{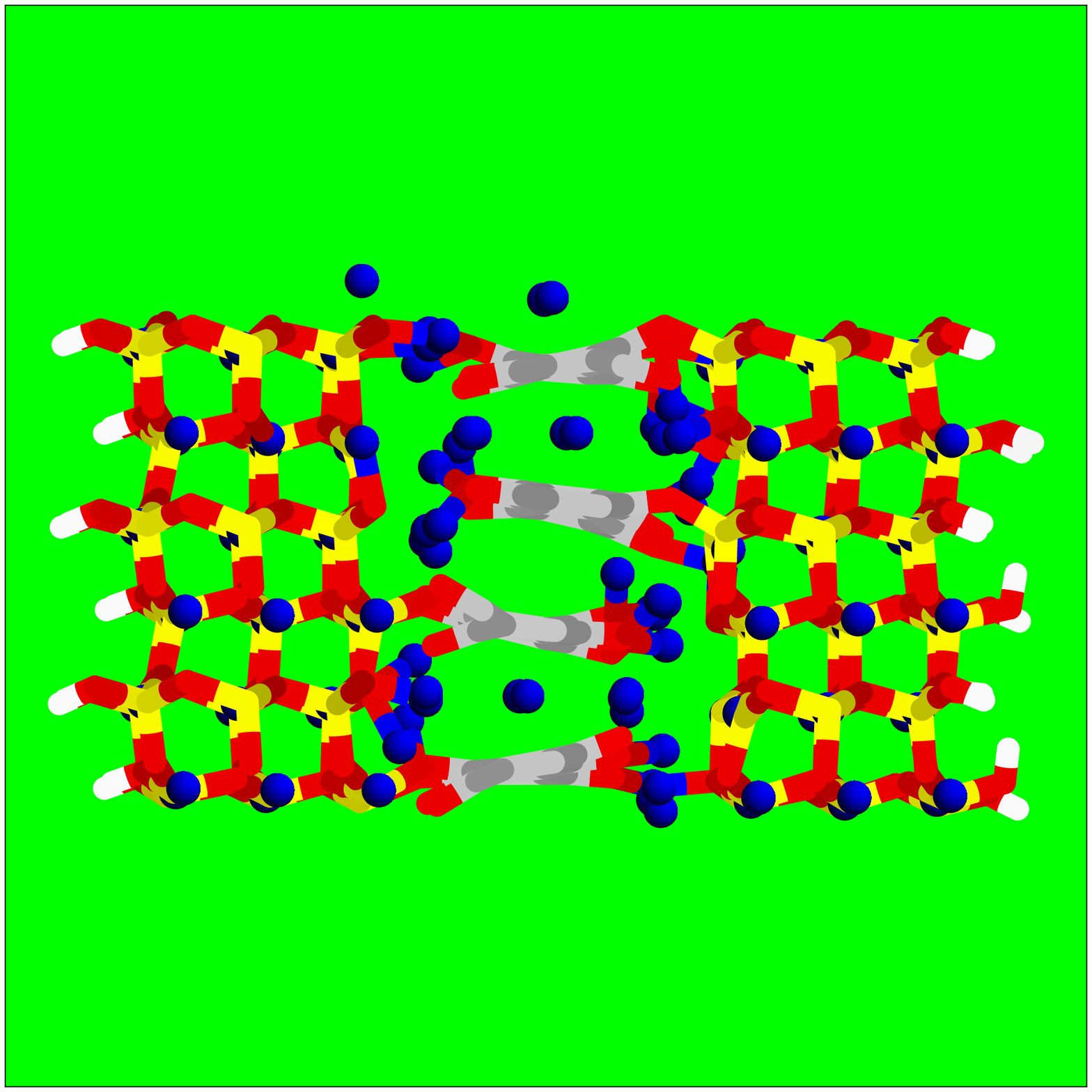}
                   (b) \epsfxsize=2.00in \epsfbox{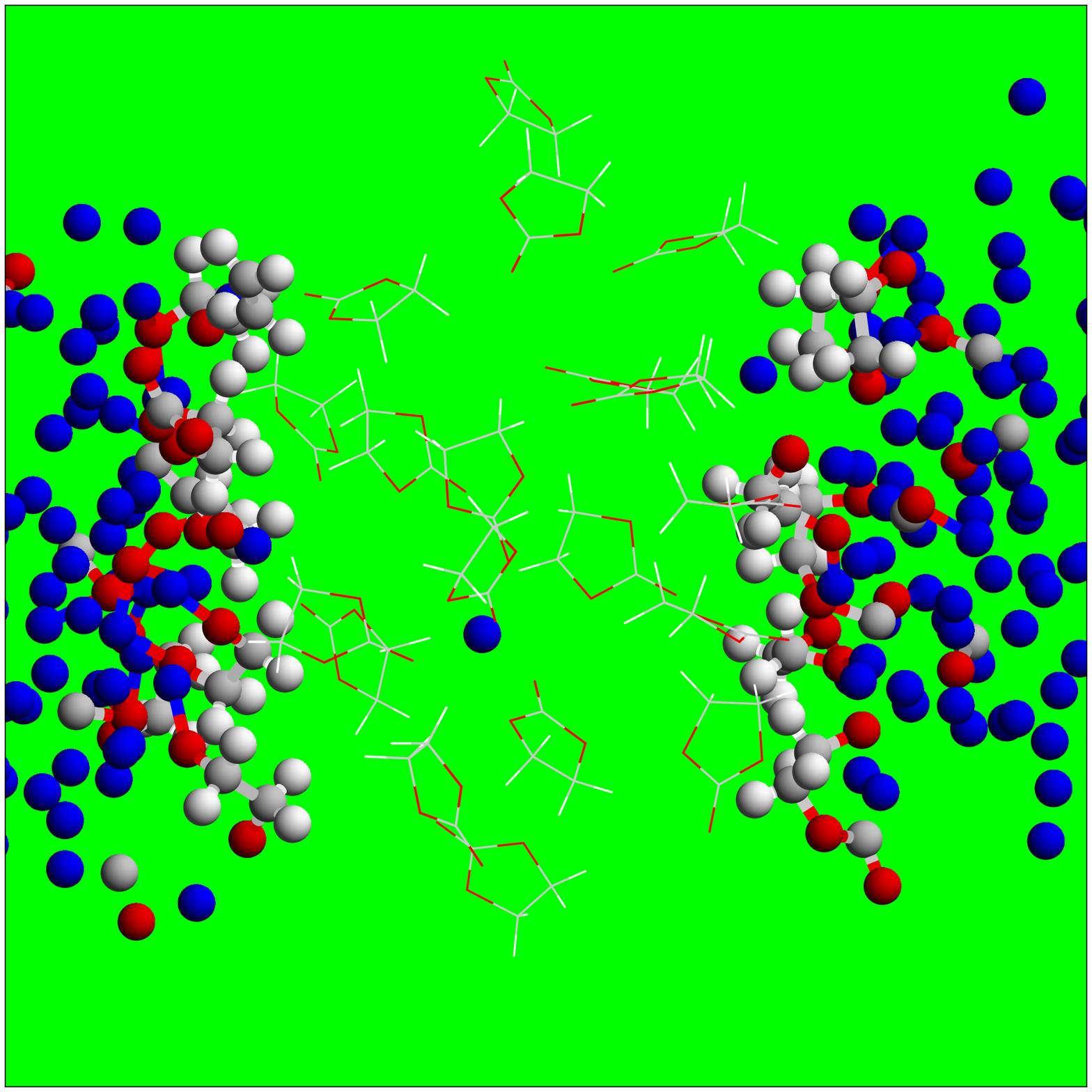}
                   (c) \epsfxsize=2.00in \epsfbox{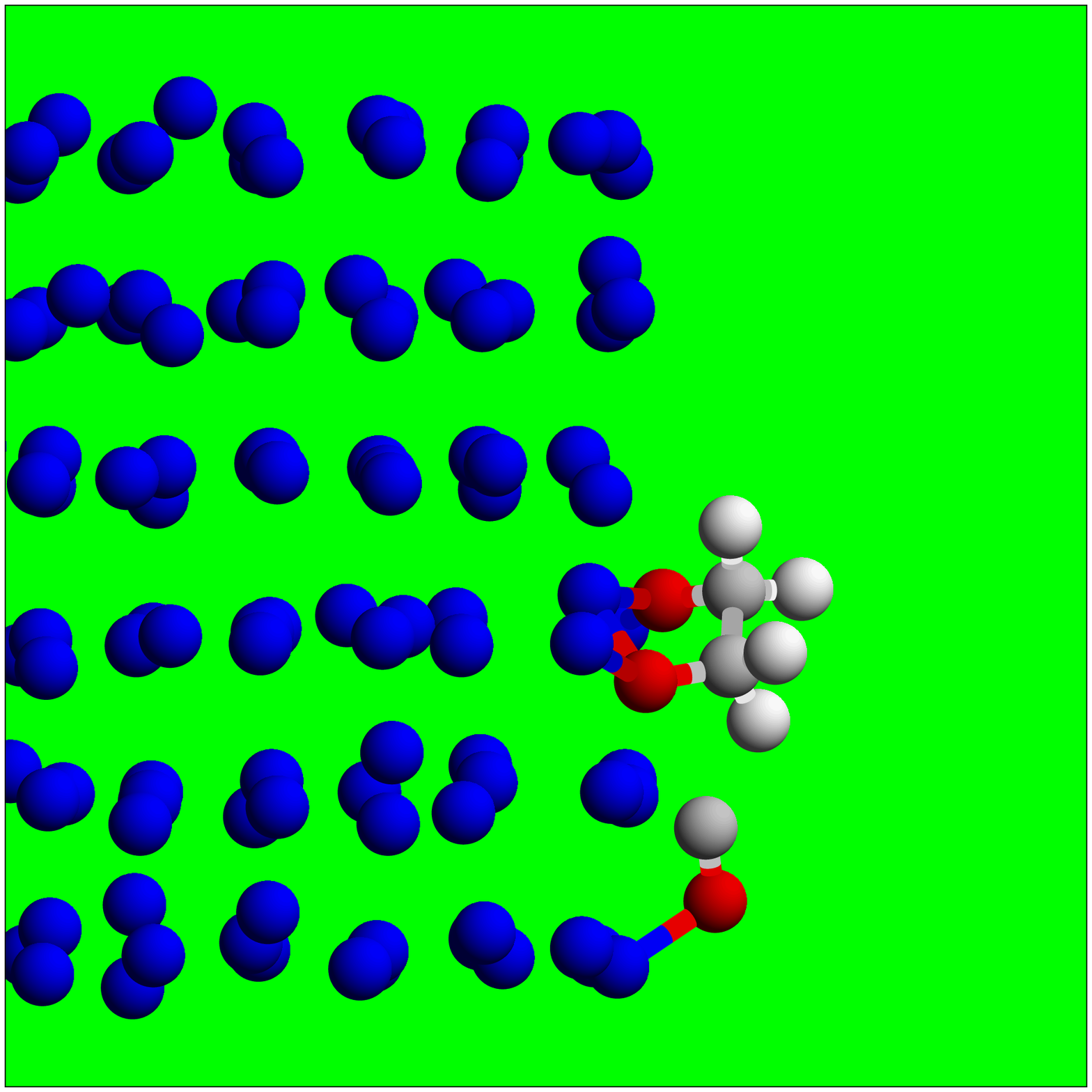} }}
\centerline{\hbox{ (d) \epsfxsize=2.00in \epsfbox{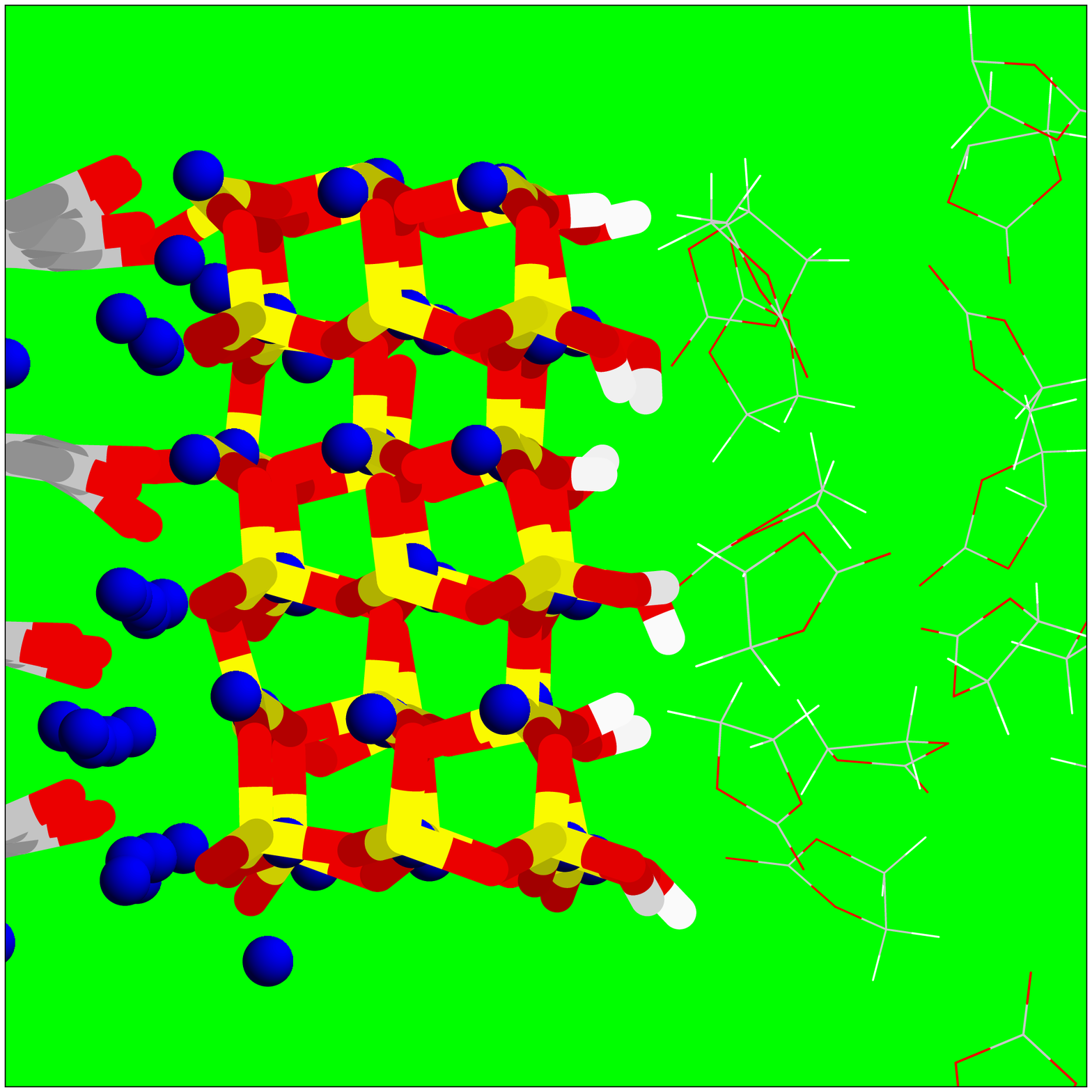}
                   (e) \epsfxsize=2.00in \epsfbox{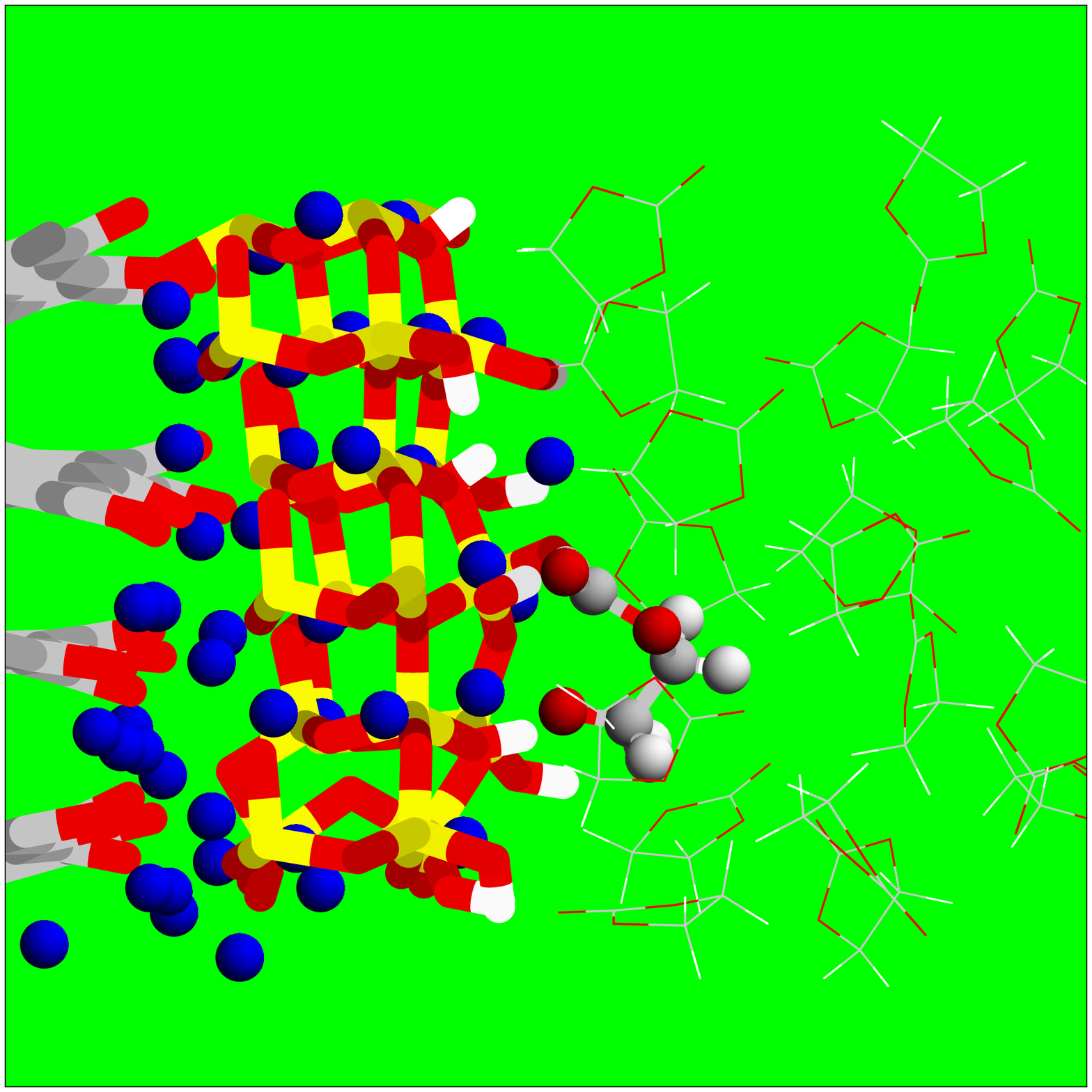} 
                   (f) \epsfxsize=2.00in \epsfbox{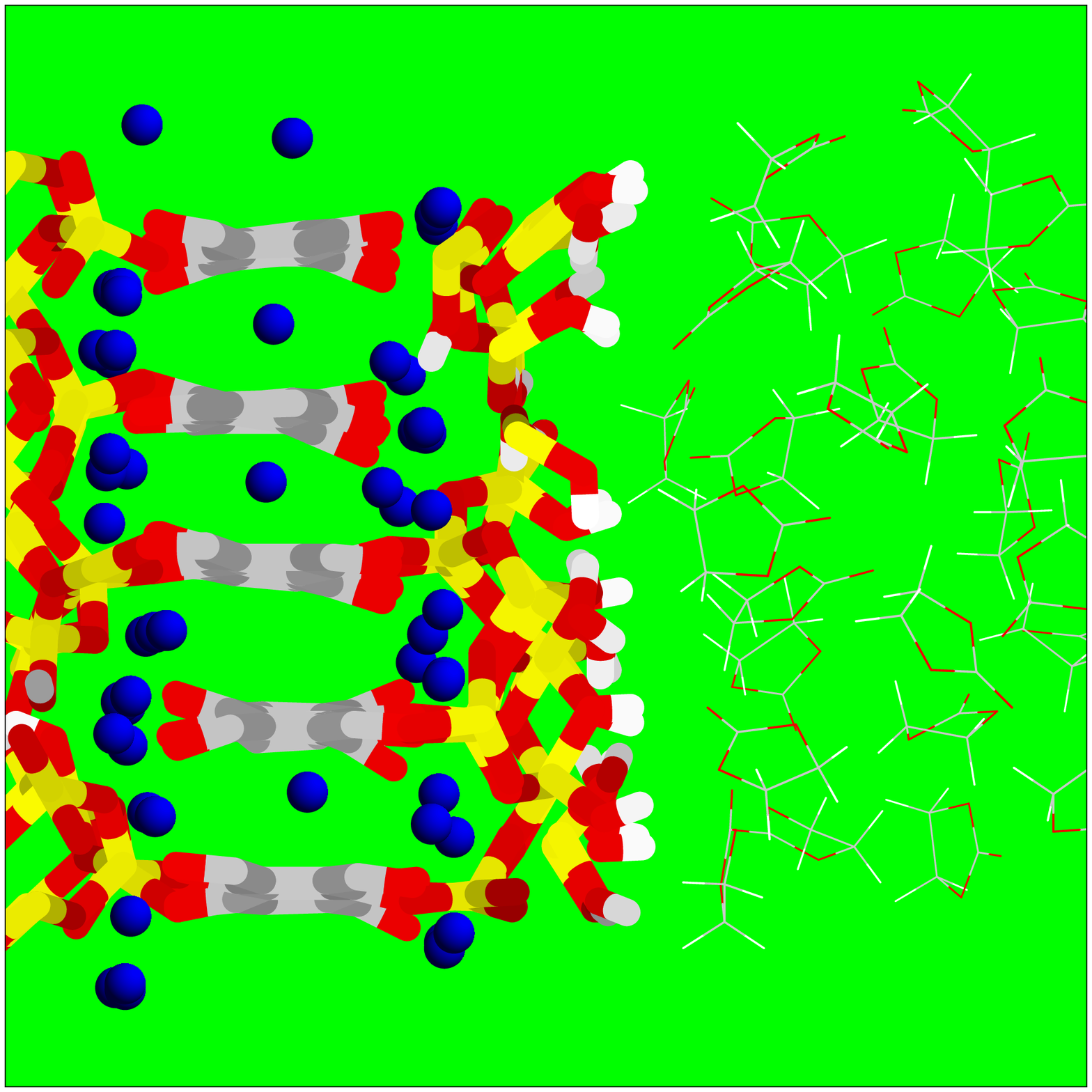} }}
\caption[]
{\label{fig2} \noindent
(a) Model electrode system with a narrow LiC$_6$ strip decorated
with C=O edges and coated with a 10~\AA\, thick LiAlO$_2$ layers terminated
by hydroxyl groups.  (b) A snapshot 15~ps into an AIMD trajectory of
32 EC molecules confined between Li metal (100) slabs conducted at T=350~K.
Li and decomposed EC are depicted as ball-and-stick models, intact
EC as wireframes.  (c) An isolated, decomposed EC on Li surface at T=350~K,
after a 7~ps AIMD simulation.  (d) No decomposition on 10~\AA\, thick
LiAlO$_2$-coating after 7~ps.  (e) One EC
decomposed on 7~\AA\, thick LiAlO$_2$-coated surface, T=450~K, after 7~ps.  
(f) No decomposition on 5~\AA\, thick, hydroxylated Al$_2$O$_3$-coating after
7~ps.  Yellow, grey, red, blue, and white depict Al, C, O, Li, H atoms,
respectively.  
%(f) Same as (d), but with only one isolated EC (no liquid)
%and a 0.4~V/\AA\, electric field, T=450~K, after 0.4~ps.  
}
\end{figure}

\begin{figure}
\centerline{\hbox{ \hspace*{0.11in} \epsfxsize=1.80in \hspace*{0.12in}
                \epsfbox{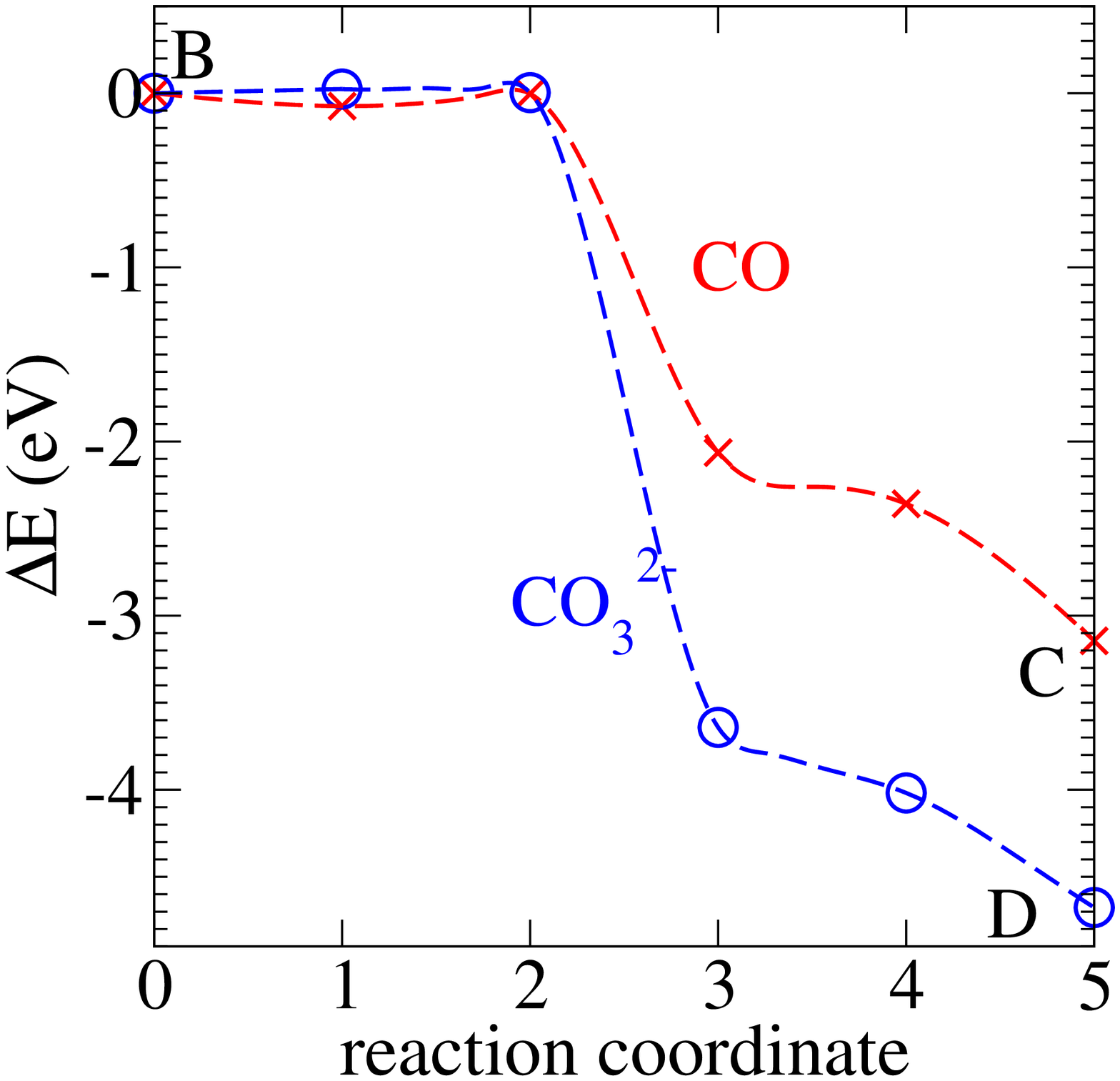}
                   (b) \epsfxsize=2.00in \epsfbox{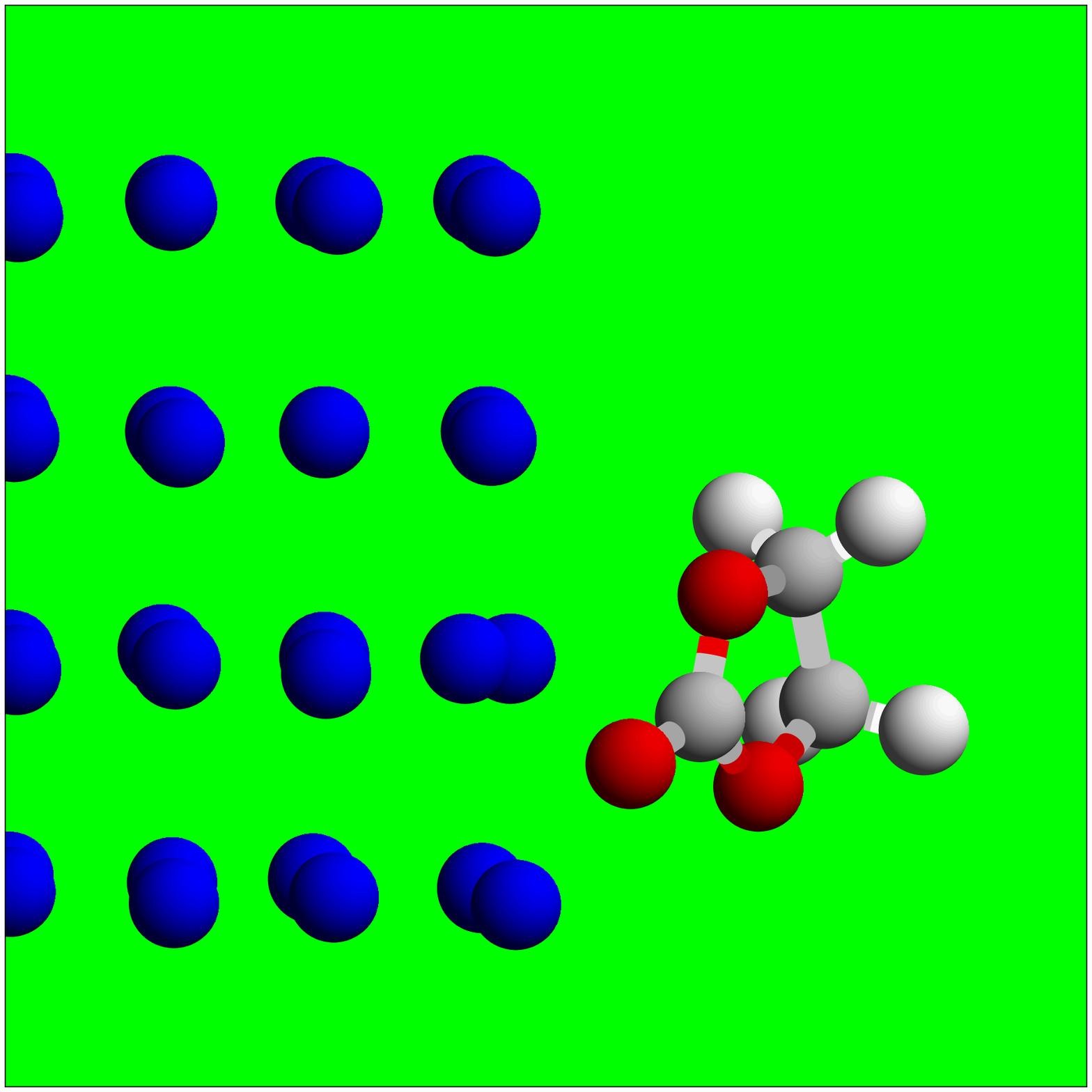} }}
\centerline{\hbox{ (c) \epsfxsize=2.00in \epsfbox{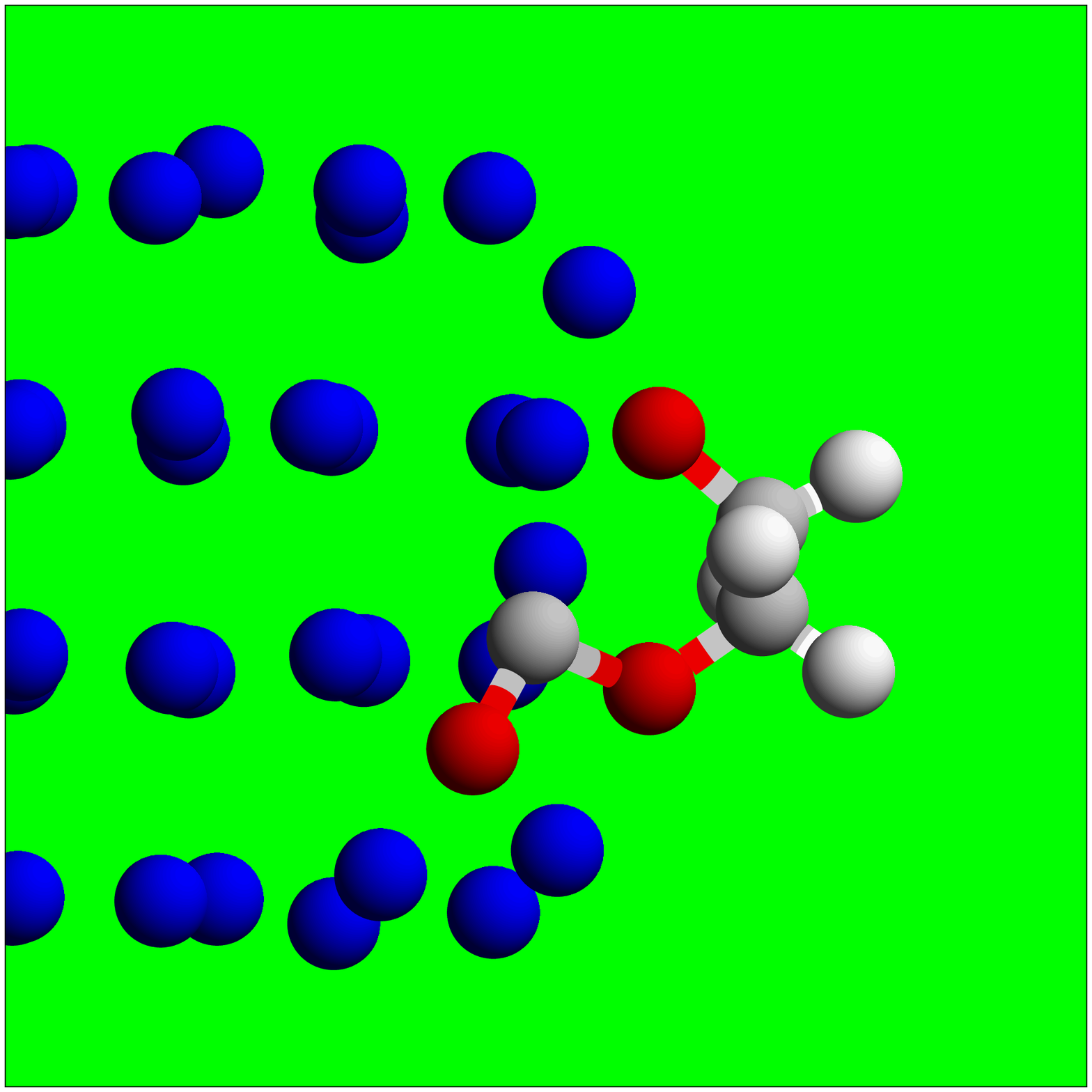}
                   (d) \epsfxsize=2.00in \epsfbox{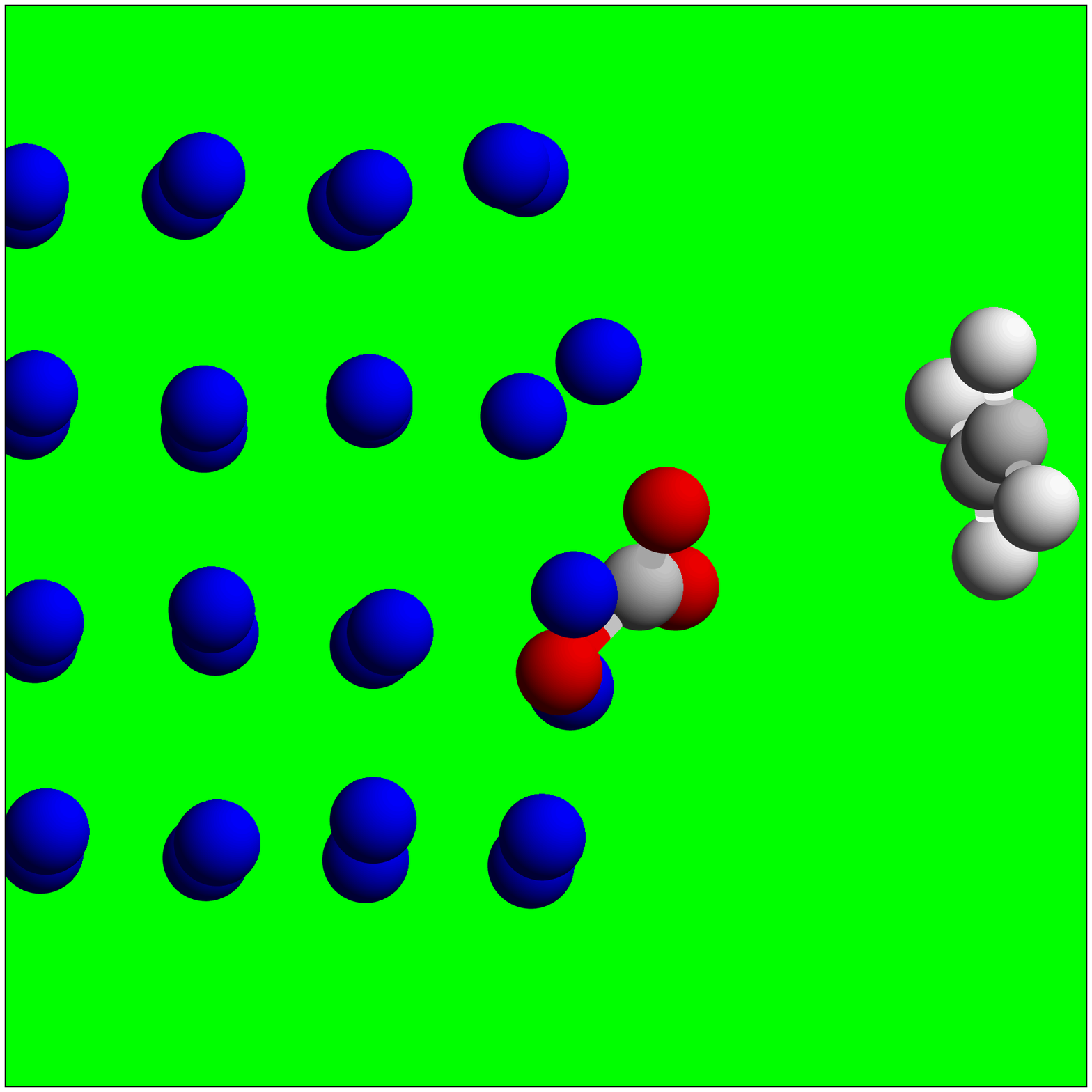} }}
\caption[]
{\label{fig3} \noindent
(a) Static, nudged elastic band (NEB) calculations of energy barriers
associated with 2 modes of EC breakdown, producing CO or CO$_3^{2-}$,
at T=0~K.  Points B, C, \& D correspond to panels (b)-(d).
(b) Intact EC on Li (100).  (c) EC partially decomposed into
OCOC$_2$H$_4$O, precursor to CO and OC$_2$H$_4$O$^{2-}$,
on Li (100).  (d) CO$_3^{2-}$ and C$_2$H$_4$ products
on Li metal.  The color key is the same as in Fig.~\ref{fig2}.
}
\end{figure}

\begin{figure}
\centerline{\hbox{ \hspace*{0.1in} \epsfxsize=2.50in \epsfbox{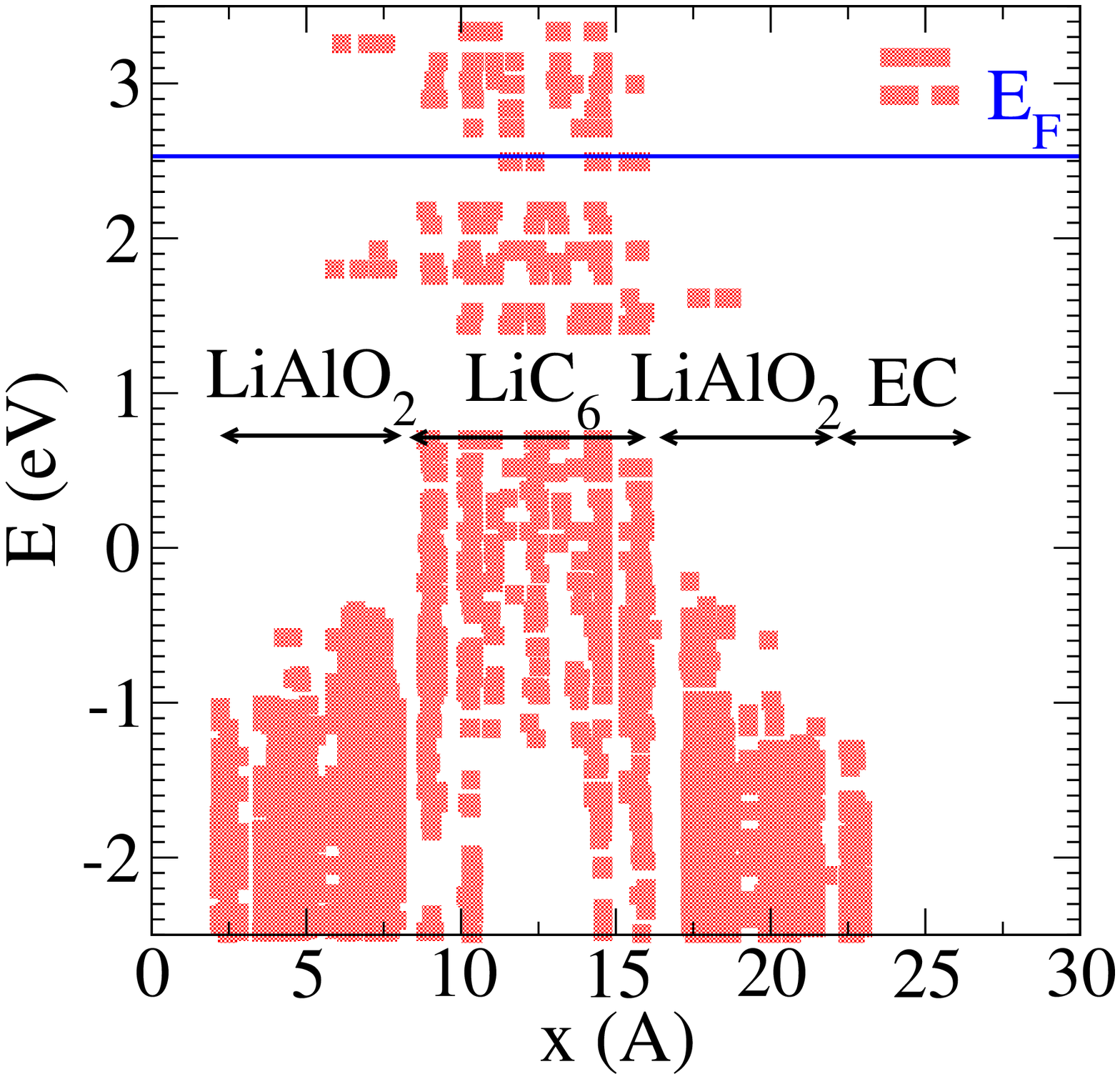}
                   \hspace*{0.1in} \epsfxsize=2.50in \epsfbox{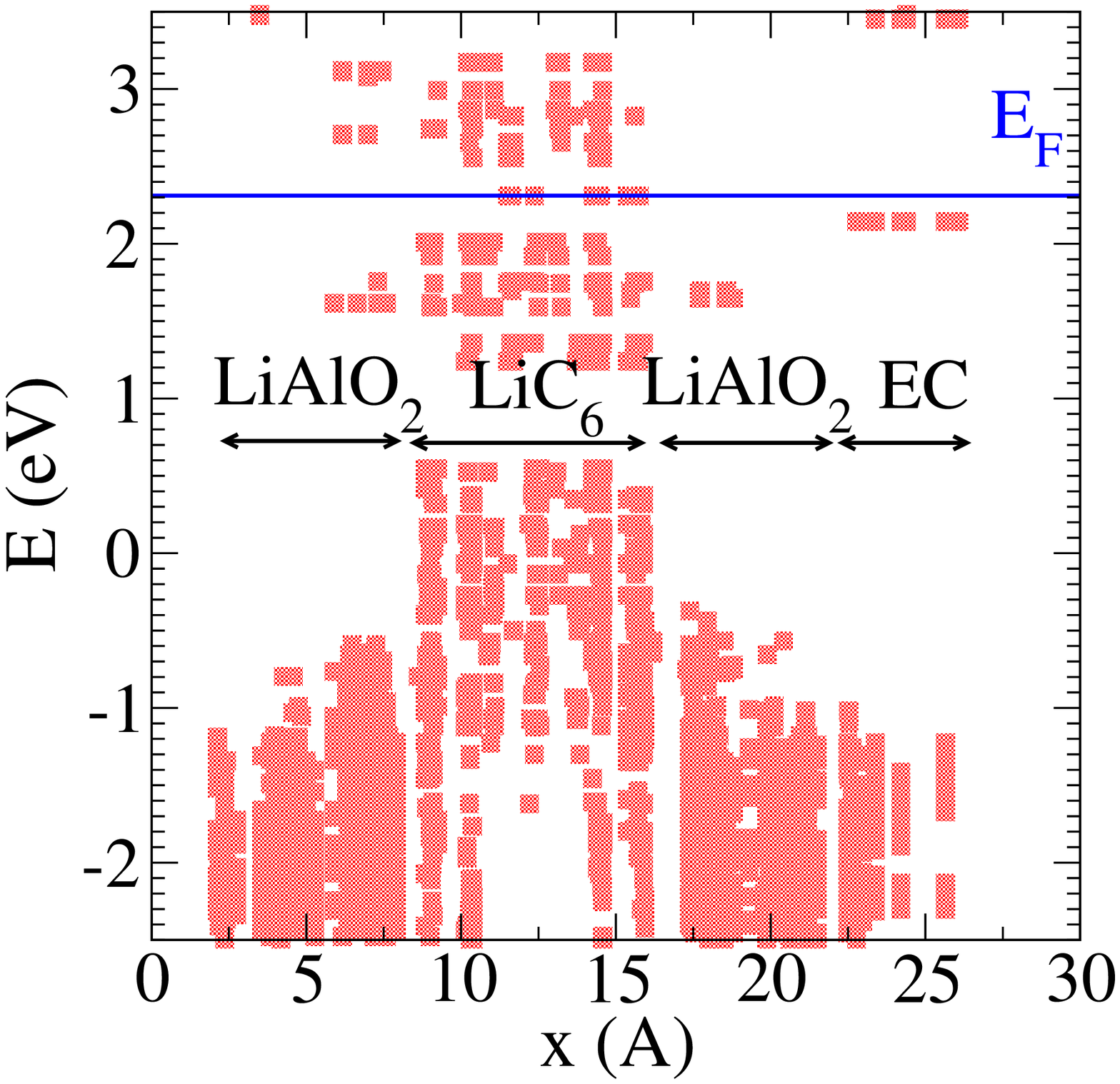} }}
\centerline{\hbox{ (c) \epsfxsize=2.50in \epsfbox{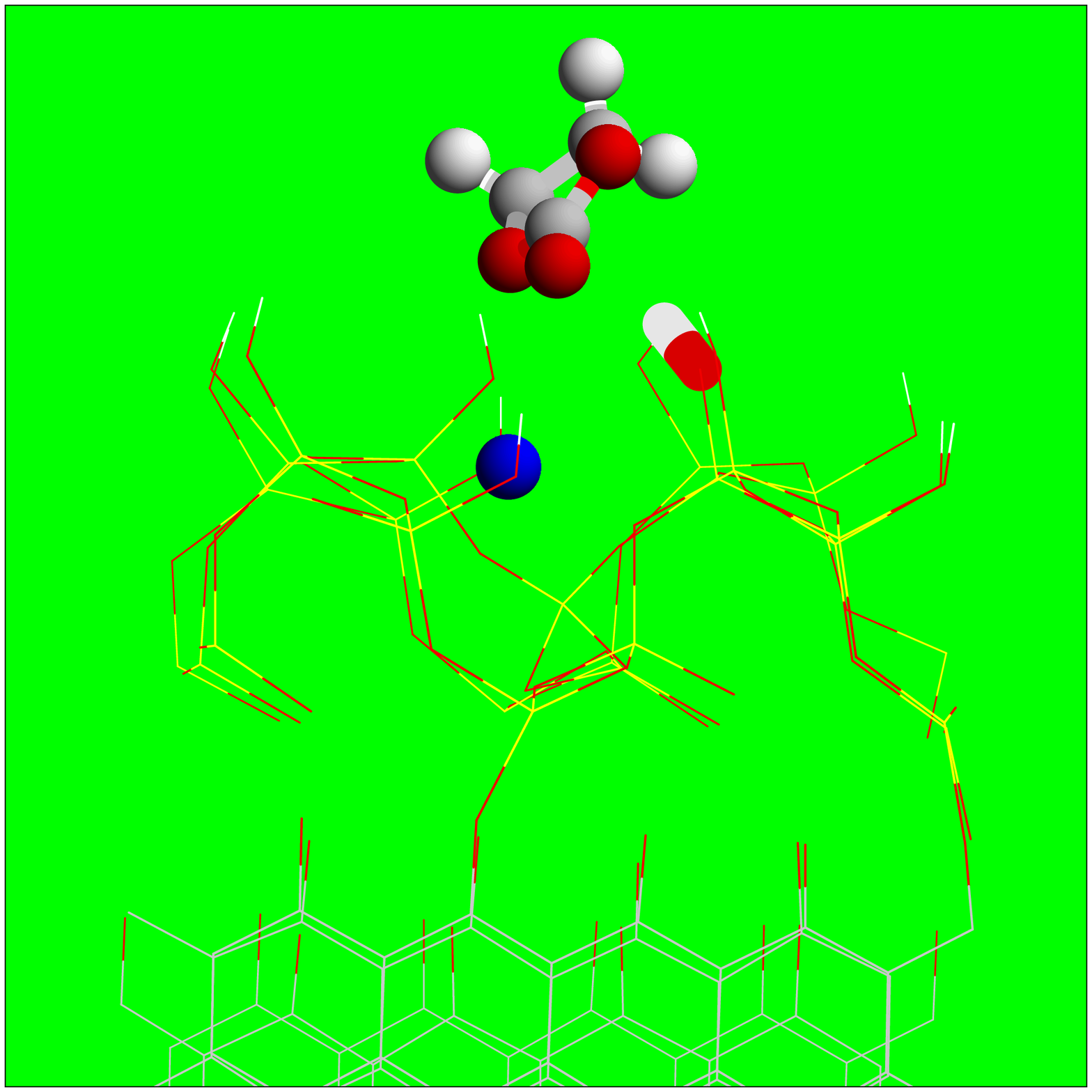}
                   (d) \epsfxsize=2.50in \epsfbox{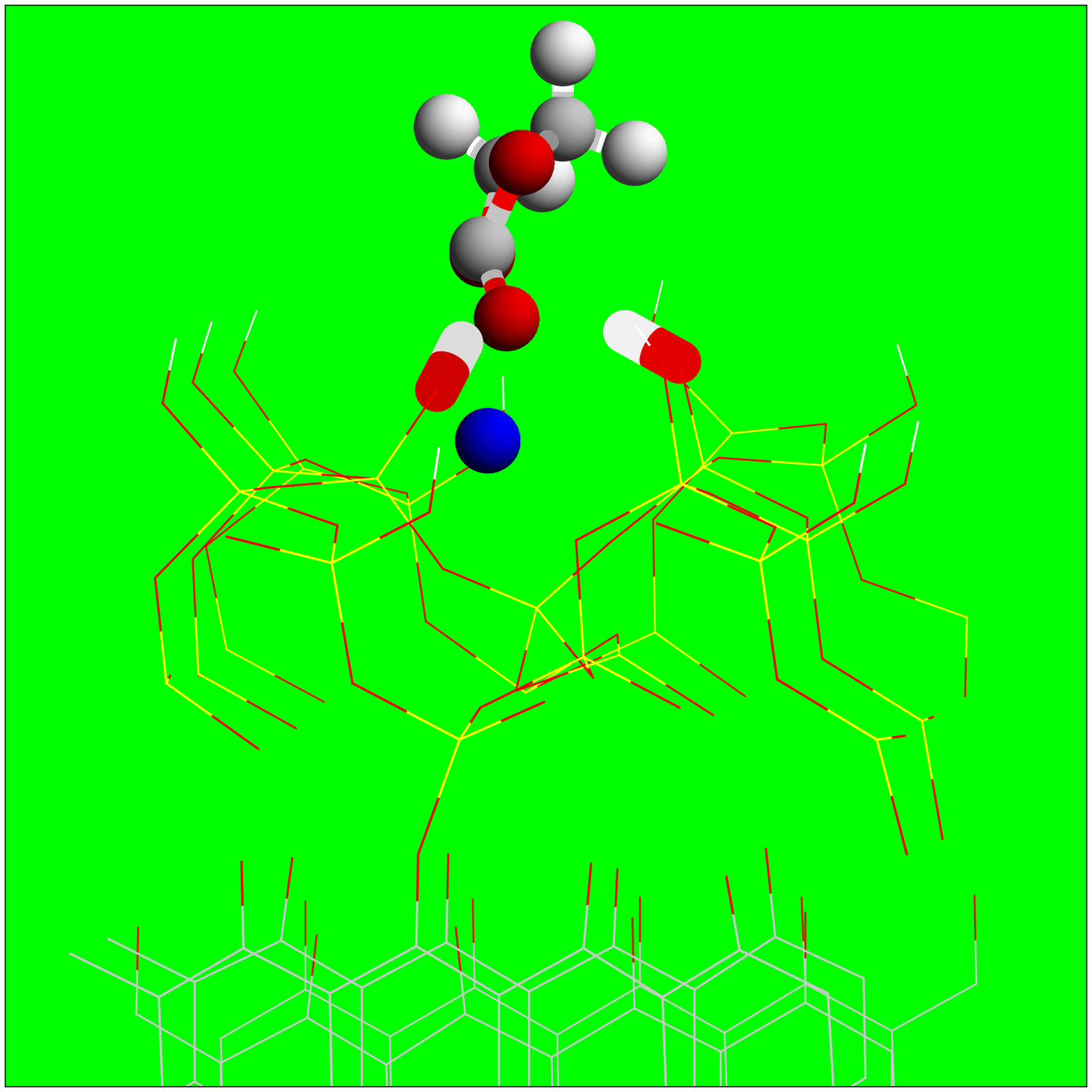} }}
\caption[]
{\label{fig4} \noindent (a)\&(b): Local electronic densities-of-state
decomposed along the $x$-axis (perpendicular to interface) for EC adsorbed
on thin LiAlO$_2$-coated LiC$_6$.
Panels (a) and (b) correspond to the configurations depicted
in panels (c) (flat EC geometry) \&(d) (bent geometry), respectively.
The red patches depict integrated up- and down-spin densities exceeding
0.01~$|e|$ for each planewave wavefunction collapsed on an atom centered
at $x$.  Panel (b) shows that the bent geometry drastically changes the
HOMO and LUMO levels, with an excess electron now residing on EC below
the Fermi level ($E_{\rm F}$).  The conduction band of the LiAlO$_2$ region
is located above $4$~eV.  In panels (c) \& (d), the EC molecule and Li
coordinated to the EC are depicted as spheres while surface hydroxyl groups
donating hydrogen bonds to the EC are stick figures.  Other Li are omitted
and all other oxide-coating and graphite atoms appear as wireframes.  EC
configurations on the thick LiAlO$_2$ coating are qualitatively similar
(not shown).  The color scheme is as in Fig.~\ref{fig2}.
}
\end{figure}

\begin{figure}
\centerline{\hbox{ \epsfxsize=6.00in \epsfbox{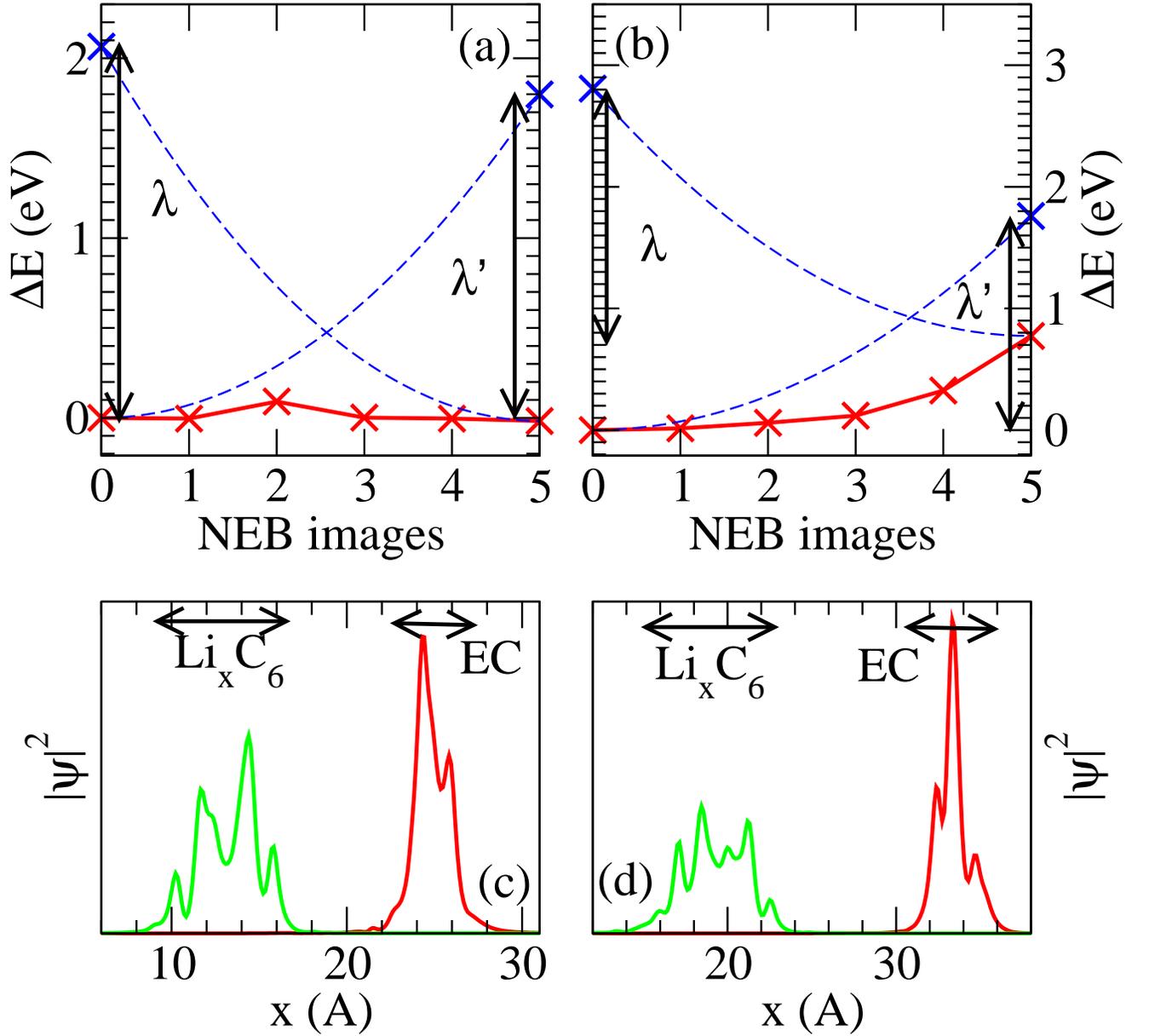} }}
\caption[]
{\label{fig5} \noindent
(a)\&(b) Adiabatic (red crosses) and non-adiabatic (blue crosses) energy
profiles along
the reaction coordinate between the flat (Fig.~\ref{fig4}c, image ``0'') and
bent (Fig.~\ref{fig4}d, image ``5'') EC geometries when applying a 0.4~V/\AA\,
applied electric field to the 7~\AA\, and 10~\AA\,-thick LiAl$_2$O layers.
Adiabatic energies are computed along the NEB-generated chain with
unconstrained DFT.  Non-adiabatic reorganization energies ($\lambda$) 
derive from cDFT.  The dashed curves are parabolic fits.  (c)\&(d) Highest
occupied orbital of the system (green), and the cDFT-computed $e^-$-accepting
EC orbital (red) adsorbed on the 7~\AA\, and 10~\AA\,-thick LiAl$_2$O layers,
integrated over the lateral dimensions.
}
\end{figure}

On the 10~\AA\, thick LiAlO$_2$-coated Li$_x$C$_6$ strip (Fig.~\ref{fig2}f),
no EC decomposes within 7~ps.  The limited duration of the AIMD trajectory
does not permit an estimate of the adiabatic AIMD/PBE free energy barrier.
While this barrier can be computed using the AIMD/potential-of-mean-force
method,\cite{silica} it will be underestimated due to PBE
self-interaction errors and underestimation of the electron tunneling barrier.

\begin{figure}
%\centerline{\hbox{ \hspace*{0.11in} \epsfxsize=1.80in \hspace*{0.12in}
%                   (b) \epsfxsize=2.00in \epsfbox{fig6b.ps} }}
\centerline{\hbox{  \epsfxsize=4.20in \epsfbox{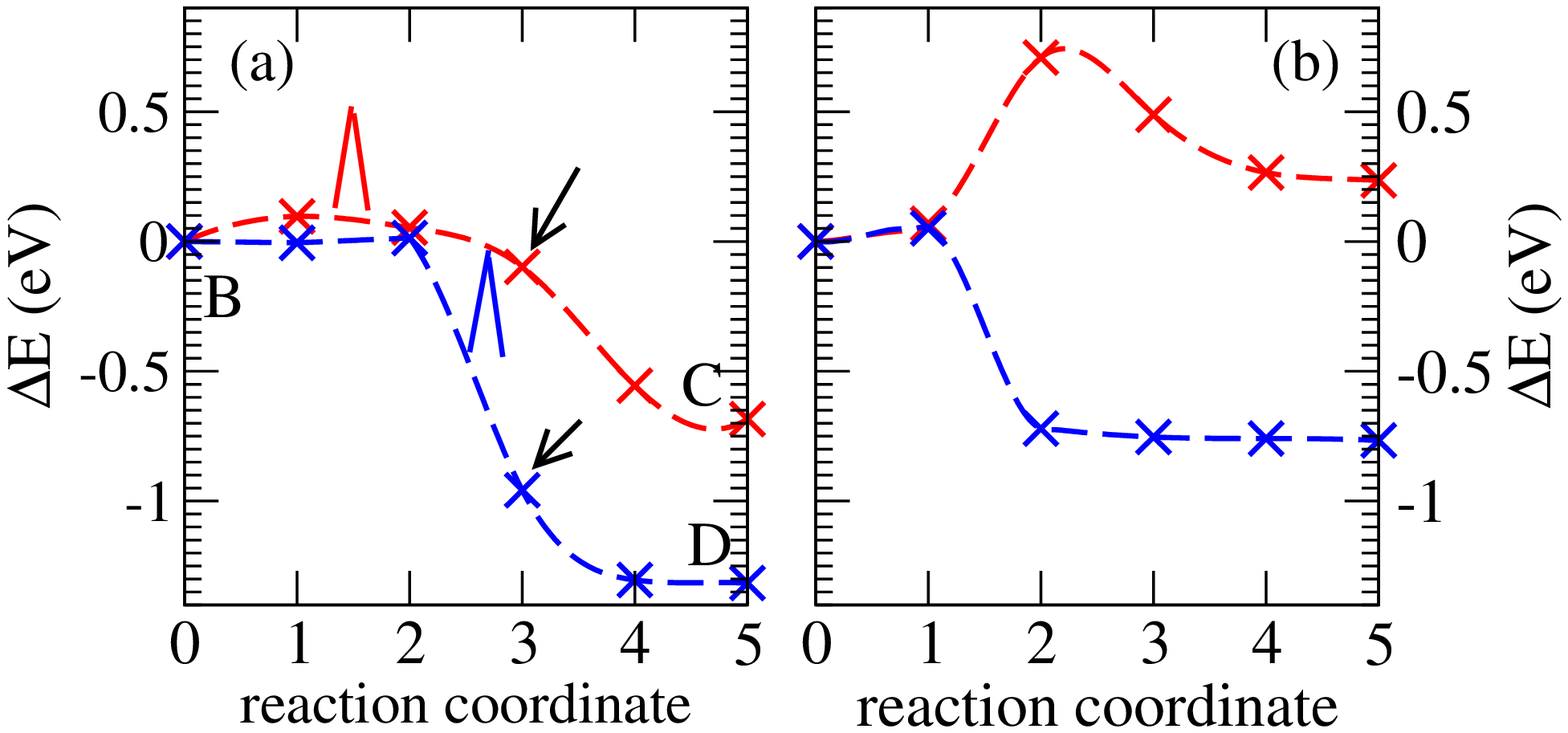} }}
\centerline{\hbox{ (c) \epsfxsize=2.00in \epsfbox{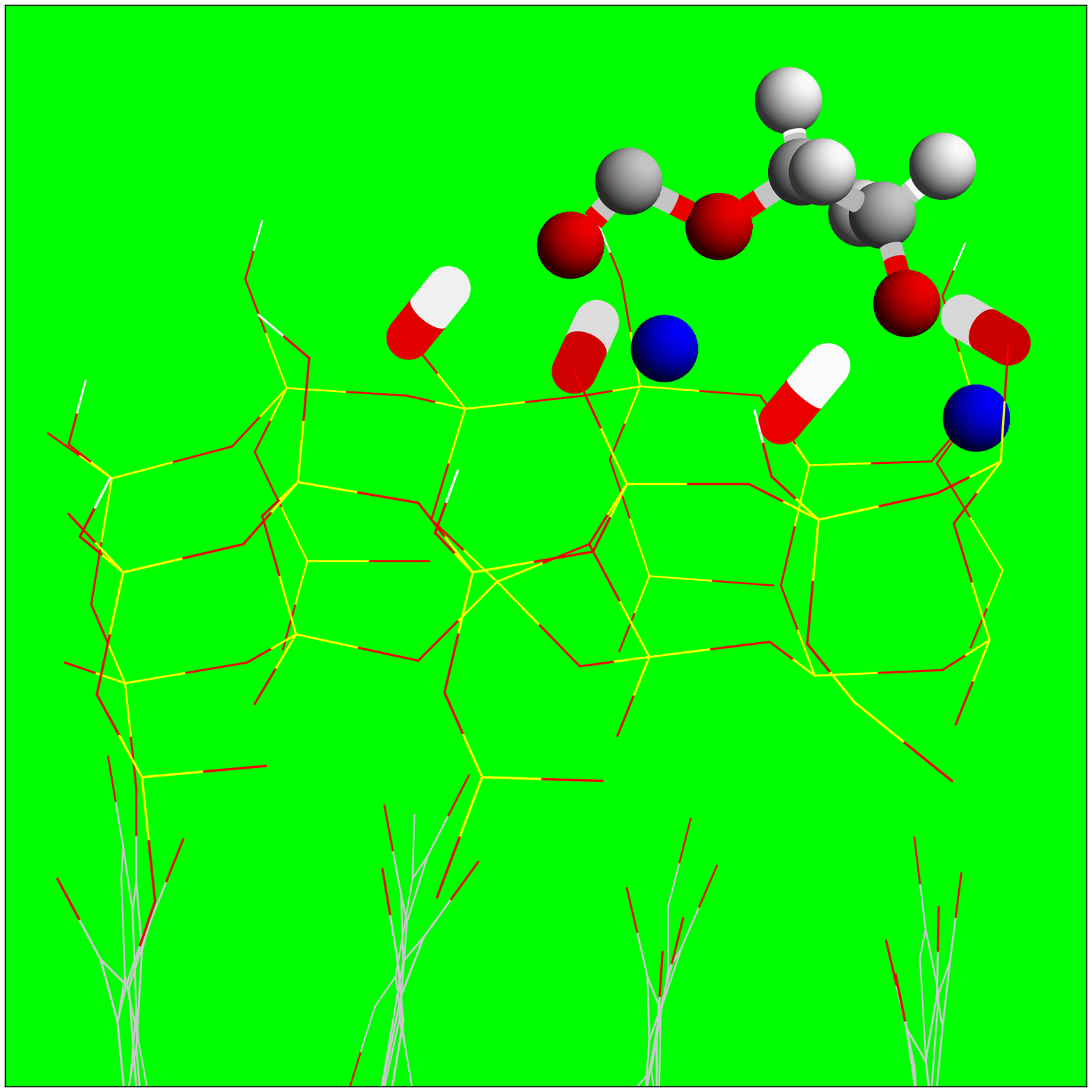}
                   (d) \epsfxsize=2.00in \epsfbox{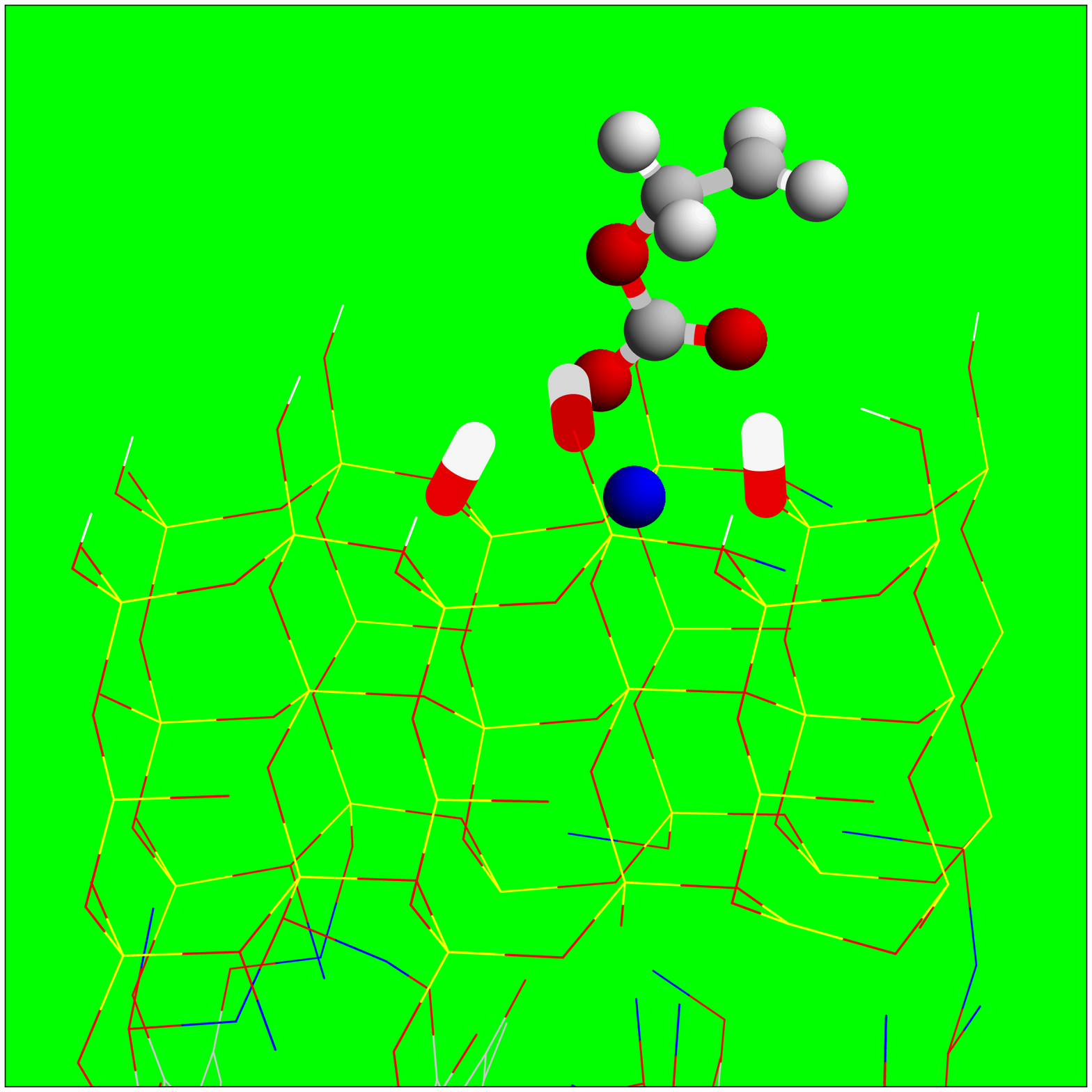} }}
\caption[]
{\label{fig6} \noindent
(a) Adiabatic DFT/PBE energy profiles associated with EC decomposition
on the thin LiAlO$_2$ coating at T=0~K.  Red and blue refer to
adsorbed OCOC$_2$H$_4$O$^{2-}$ and C$_2$H$_4$OCO$_2^{2-}$ intermediates,
which are precursors to CO and CO$_3^{2-}$ products, respectively.
The electric field strength is
0.4~V/\AA.  The dashed lines are guides to the eye.  Red/blue triangles
depict the barrier associated with the first transfer of an electron
on to the EC molecule, which is detected between images.  They indicate
that the DFT/PBE treatment erroneously neglects the electron tunneling
barrier.  Black arrows denote first detection of C-O bond breaking in an
image.  Point~B correspond to the intact EC in Fig.~\ref{fig4}c. (b)
Same as (a) but for a single 10~\AA\, thick layer in the absence
of Li$_x$C$_6$ or electric field.
(c)-(d) OCOC$_2$H$_4$O$^{2-}$, and C$_2$H$_4$OCO$_2^{2-}$ on
thin LiAlO$_2$ surfaces, corresponding to points C~\&~D in panel (a).  
The atom representation is as in Fig.~\ref{fig4}.
}
\end{figure}

However, on the 7~\AA\, thick LiAlO2$_2$ layer, a C$_{\rm C}$-O bond
on one EC molecule is spontaneously broken within 1~ps (Fig.~\ref{fig2}e),
yielding OCOC$_2$H$_4$O$^{-}$, the majority predicted product on Li metal
surfaces (Fig.~\ref{fig2}b) and a precursor to CO.  Here the monovalent anion
intermediate is stabilized by hydrogen bond donation from several AlOH
groups and by coordination to two surface Li atoms.  Since the DFT/PBE method
erroneously underestimates the 0.51~eV $e^-$ tunneling barrier associated
with molecular reorganization (Fig.~\ref{fig4}a) which precedes bond-breaking,
it {\it vastly overestimates the overall bond-breaking rate}.
Indeed, the EC decomposition timescale predicted with DFT/PBE is similar
similar to the timescale predicted in the {\it absence} of the ALD
layer.\cite{ec}  This is in disagreement with our experimental measurements 
which reveals far less solvent decomposition products when an ALD layer is
present (Sec.~\ref{expt}).  Instead, 0.51~eV should be taken as the overall
activation energy in these bond-breaking events.  With this barrier, the
bond-breaking rate should occur in millisecond, not picosecond, timescales
at room temperature.  Nevertheless, this PBE-based AIMD calculation is
valuable because it identifies the most reactive
surface site.  An EC adsorbed at this site is used in the $e^-$ transfer
calculation of the previous section (Fig.~\ref{fig4}).  Under UHV-like
conditions, an isolated EC molecule adsorbed at this site exhibits $<0.05$~eV
adiabatic DFT/PBE C-O bond-breaking barriers provided a 0.4~V/\AA\,
electric field is applied (Fig.~\ref{fig6}).  The qualitative correspondence
between adiabatic AIMD/PBE decomposition rate and UHV barrierless reaction
is the reason this model is adopted for $e^-$ transfer studies in
Sec.~\ref{nonadiab1}.

Because of its extreme thinness, optimizing the 7~\AA\,-thick LiAlO$_2$
film coated on to Li$_x$C$_6$ has caused 2~Li atoms per surface to migrate
outwards (Fig.~\ref{fig2}e).
These outlying Li coordinate to the surface hydroxyl groups, polarizing
them.  The EC that undergoes breakdown (Fig.~\ref{fig6}d) is indeed hydrogen
bonded to an OH group coordinated to a surface Li$^+$.  Such
Li migration to the surface does not occur in the thicker LiAlO$_2$ coating.  
Hence the faster adiabatic AIMD/PBE EC decomposition dynamics on the thin
LiAlO$_2$ coating is not just a consequence of oxide thickness, but
is partly due to active site chemical specificity.  This anomaly may
also be the reason the predicted $V_{\rm AB}$ value does not strongly
decrease with increasing the oxide thickness from 7~\AA\, to 10~\AA,
and may further explain the difference in work functions between Li$_x$C$_6$
coated with 7~\AA\, and 10~\AA\, thick LiAlO$_2$ films (Table~\ref{table2}).

The 10~\AA\,-thick LiAlO$_2$ coating does not exhibit outward Li atom
migration.  Here the DFT/PBE bond-breaking barriers of adsorbed EC are not 
readily deconvolved from $e^-$ transfer (S.I.).  For simplicity, we consider
a model with just one 10~\AA\, thick LiAlO$_2$ layer hydroxylated on both sides
(Table~\ref{table1}), add one excess $e^-$ that now {\it always} resides
on the EC because of the Li$_x$C$_6$ $e^-$ sink has been removed, and
compute EC$^-$ decomposition energetics without applied electric fields.
C$_{\rm E}$-O bond-breaking to form CO$_3^{2-}$ precurors remain
barrierless and exothermic.  However, the C$_{\rm C}$-O cleavage route to
form CO precurors becomes endothermic and exhibits a 0.71~eV barrier.
This indicates a product channel cross-over as the oxide thickness
increases and/or the reactivity of the surface site decreases.  The expected
reaction pathyways transition from a mixture of C$_{\rm E}$-O and C$_{\rm C}$-O
bond breaking to predominantly C$_{\rm E}$-O cleavage (CO$_3^{2-}$ precursor).
While the liquid solvent environment is not included here, we speculate
that this finding may be extrapolated to other coating surfaces, including
natural SEI films, as the surface sites become less reactive.  In the
future, we will also examine EC decomposition reactions on Li$_2$CO$_3$
surfaces to see if similar trends persist on that crystalline
material, recently adopted as a theoretical model for organic solvent
decomposition SEI film, and the decomposition of other solvent/salt
molecules.\cite{iddir}

We have also conducted AIMD simulations of graphitic anodes coated
with 5~\AA\, thick hydroxylated Al$_2$O$_3$ layers (Fig.~\ref{fig2}f).
No Li ions reside near the interface region, and no solvent decomposition
is observed within 7~ps, despite the thinness of the oxide.  This
emphasizes the importance of surface heterogeneity at atomic lengthscales.
Replacing all surface AlOH groups with AlOLi dramatically increases
the decomposition rate; this will be discussed in future publications.

\section{Experimental Results}

\label{expt}

\begin{figure}
\centerline{\hbox{ \epsfxsize=5.50in \epsfbox{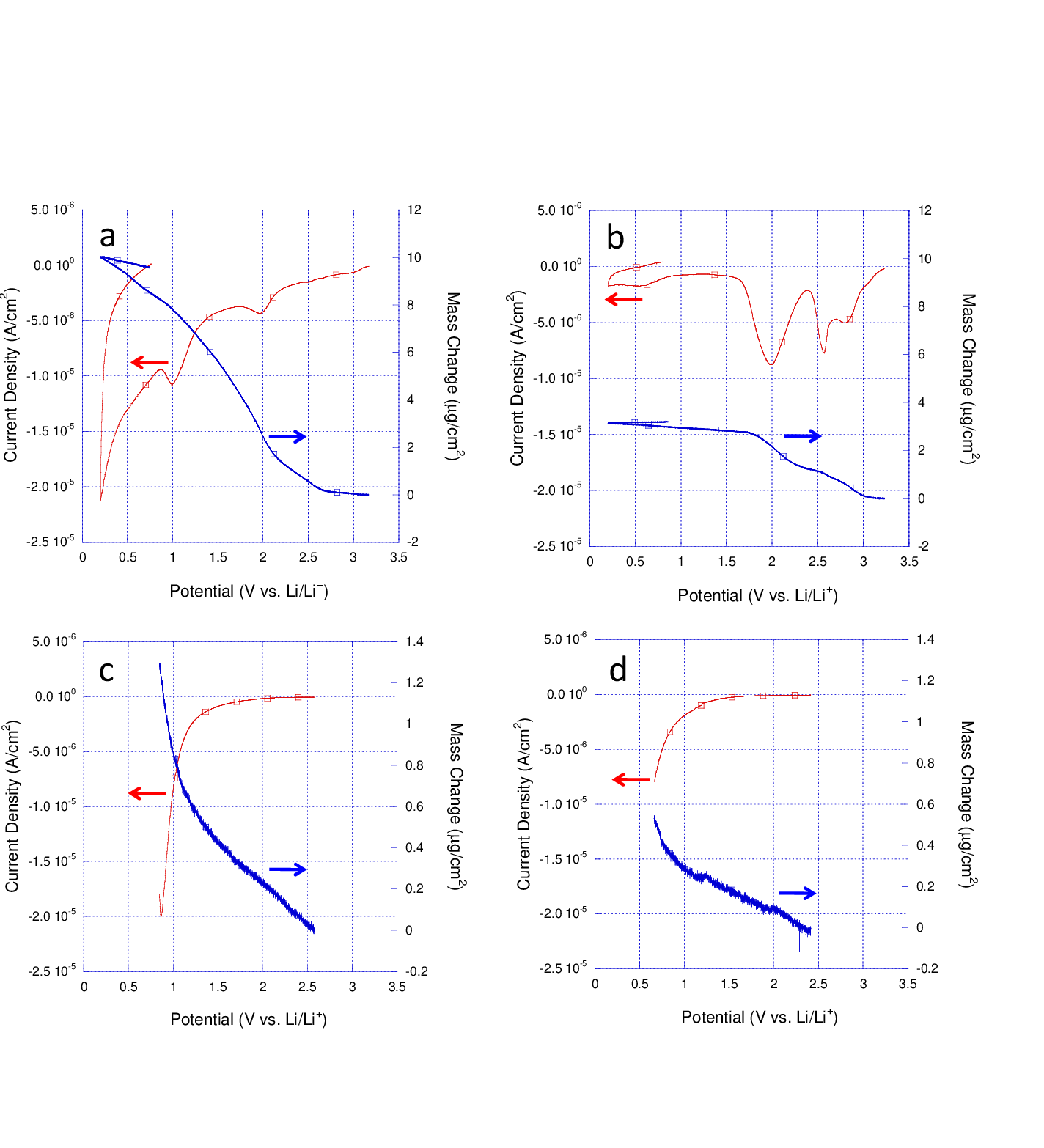} }}
\caption[]
{\label{fig8} \noindent
Current and mass change response with 1~mVs$^{-1}$ cathodic
polarization of an electrode in 1~M LiPF6, 1:1~vol.~EC:DEC.
(a) 50~nm C~film, (b) Cu~substrate without a C~film, (c) 0.55~nm thick
ALD Al$_2$O$_3$ on a 50~nm C~film, and (d) 1.1~nm thick ALD Al$_2$O$_3$ on
a 50~nm C~film. Current and mass are normalized to the geometric area.
}
\end{figure}

Figure~\ref{fig8} shows the combined voltammetric and microgravimetric
responses of the uncoated and alumina coated PLD carbon films as the electrode
potential is decreased to a value slightly above the threshold for Li$^+$
intercalation in the carbon. The uncoated carbon electrode (Fig.~\ref{fig8}a)
exhibits a continuously increasing current response, with several discrete
maxima.  One maximum reaches a value of 4~$\mu$A/cm$^2$ with a mass increase of
2~$\mu$g/cm$^2$ at a potential of 2~V.  The
other maximum reaches 11~$\mu$/cm$^2$ at a potential of 1~V.  The
decomposition of the electrolyte and deposition of byproducts at 2~V
is catalyzed by the Cu substrate, as evidenced by the
similar current and mass changes on a control Cu electrode (Fig.~\ref{fig8}b),
and demonstrate that the carbon films possess porosity and allow electrolyte
penetration. As seen in the limiting current and mass profiles of
Fig.~\ref{fig8}b,
electrolyte decomposition results in Cu passivation beyond 2~V, arguing
that the majority of the current and mass changes measured above 2~V for the
porous carbon films (Fig.~\ref{fig8}a) are due to electrolyte decomposition
on Cu; only the signal below 1~V is associated with
solid electrolyte interphase formation (SEI) on the carbon surface. The
porous and therefore higher area carbon surface exhibits a continuous
increase in both current and mass uptake as the potential is further
reduced from 2 to 0.2~V and the onset of Li$^+$ intercalation is approached.
With an approach to 0.2~V, the rate of current change increases
substantially over the rate of mass change, signaling a point where
Li$^+$ intercalation has initiated, where the lighter mass Li (compared
to an fragment of ethylene carbonate or diethyl carbonate) accounts for
a growing fraction of the measured current. The possibility exists that
current increase could also be related to solvent reduction without
mass addition to the surface (soluble byproduct formation), but note
that the mass decrease upon reversal of the potential sweep clearly
argues for the onset of Li$^+$ ion intercalation into the carbon.
We note that the scan rate of 1~mV/s is sufficiently fast to
produce only modest extraction of Li during this reverse partial
half cycle. 

The alumina coating acts as a kinetic barrier to prevent electron transfer to
the organic carbonate molecules of the electrolyte. Figures~\ref{fig8}c,~d
show the response of a 0.55 and 1.1~nm thick coated carbon films to
the onset of electrolyte reductive decomposition. Comparison of the
uncoated (Cu subtracted) and coated carbon films shows that a higher
overpotential is required to drive solvent decomposition and a
lower quantity of mass addition takes place with the alumina coating
present. A Cu current and mass uptake response is eliminated for
these coated electrodes because the alumina nucleating agent and film
precursors fully penetrate the porous carbon, conformally coating
both the carbon network and the underlying exposed regions of the
Cu substrate. The onset for significant current density and mass
increase occurs at approximately 1.2~V and 0.8~V for the 0.55~nm
and 1.1~nm alumina coatings, compared to 1.5~V for the uncoated
carbon. Mass increases measured at 0.8~V are 6, 1.3~and
0.5~$\mu$g/cm$^2$ for the uncoated (Cu subtracted), 0.55~nm and~1.1~nm
alumina sample, respectively. The greater overpotential and reduced
mass uptake of the 1.1~nm coating relative to the thinner 0.55~nm
coating argue that the thicker film provides a more effective kinetic
barrier for reducing the extent of both reductive solvent decomposition
and byproduct deposition on the electrode. The thicker alumina film
would be expected to present a lower electron tunneling rate resulting
in a slower rate of solvent decomposition and retarded SEI formation.
The fact that mass addition is observed in the presence of these
alumina coatings is a clear indicator that alumina serves to retard
and limit the extent of but does not prevent electrolyte reduction and
resulting byproduct film formation.

\section{Conclusions}

\label{conclude}

In this work, we compare EC decomposition on Li metal and on models of
oxide-coated electrodes.  The latter mimics recent experimental work using ALD
technique to passivate anodes.  This ALD strategy carries significant
technological promise,\cite{dillon,dillon0,dillon1,dillon3,dillon4,dillon2}
and it also provides an ideal robust platform for theoretical and experimental
study of passivating mechanisms.  These two systems represent two electron
transfer regimes.

On pristine Li (100) surfaces, liquid EC and even isolated adsorbed EC
molecules are predicted to undergo decomposition in picosecond time scales.
CO is the dominant product from EC, possibly because of favorable kinetic
prefactors, even though both the CO and CO$_3^{2-}$ reaction pathways
are almost barrierless and the CO$_3^{2-}$ product is more thermodynamically
stable.  EC molecules and the electrode are in close contact and
strongly coupled.  Adiabatic DFT/PBE and AIMD/PBE simulations should
be accurate in this regime.

In contrast, electron transfer through an oxide layer should be slow
compared to nuclear motion.  We find evidence that tunneling through
even a 7~\AA\, thick oxide layer belongs to the non-adiabatic regime.  
Applying constrained DFT (cDFT) calculations, such thin coatings are
found to slow down $e^-$ transfer because the solvent
reorganization energy $\lambda$ now figures prominently in electron
tunneling through the oxide.  $\lambda$, largely neglected in previous
studies of electrolyte decomposition in batteries, is estimated to be
$\sim 2$~eV for adsorbed EC molecules in ultra-high vacuum-like
conditions.  This translates into a $\sim 0.5$~eV electron tunneling
barrier within the harmonic approximation when the $e^-$ transfer
free energy change is small.

cDFT calculations show that the 7~\AA\,- and 10~\AA\,-thick LiAlO$_2$
coated Li$_x$C$_6$ exhibit electron transfer rates of $\sim 10^5$/s
at the Li$^+$/Li(s) applied voltage.  The predicted $e^-$ transfer
rate is not free of ambiguities and assumptions, and is of order-of-magnitude
utility; further fundamental research is needed for a more rigorous
treatment.  Despite this caveat, this work respresents the first
first-principles estimate of the $e^-$ tunneling rate between an
electrode and an EC molecule across an insulating oxide layer.  Such
predictions are critical for understanding ALD-hindered SEI growth in
lithium ion batteries.

The overall electron transfer rate (Eq.~\ref{nonadiab} or Eq.~\ref{et_final})
also depends on the offset $\Delta G_o$ between $e^-$ donor and acceptor
species.  $\Delta G_o$ in turn depends on the applied voltage.  AIMD estimates
of $\Delta G_o$ in an explicit liquid solvent environment is currently
lacking, and we have relied on dielectric continuum treatments of the
liquid environment.  Nevertheless, our analysis yields useful insights.
With any reasonable estimate of $\Delta G_o$, the electron
transfer rate to EC at the surface is predicted to be faster than 1/s,
and solvent breakdown on the ALD oxide is expected.  This is confirmed by
our gravimetric measurements on ALD-coated anodes, although the amount of
solvent decomposition product is significantly less than that on uncoated
graphite electrodes.  

In the case of oxide-coated electrodes, AIMD/PBE and DFT/PBE calculations
without electronic constraints vastly underestimate the electron
transfer barrier.  The reason is most likely the self-interaction error,
which unphysically favors a split electron partially localized on the
EC and partially delocalized on the electrode.  This defect exists in
many DFT functionals and has been known to yield errors in when a molecular
is split into two fragments.\cite{wtyang}  As a result, direct AIMD/PBE
simulations overestimate EC decomposition rates at oxide-coated electrode
surfaces by many orders of magnitude.  However, AIMD/PBE and DFT/PBE
calculations still provide a wealth of information about structure and
relative energetics, and they form the basis of Marcus theory considerations
and non-adiabatic electron transfer studies which are key aspects of this work.

Taking advantage of the qualitative correspondence between AIMD liquid
state reaction rates and ultra-high vacuum-like DFT calculations of
barrier heights at T=0~K in an electric field, we have applied calculations
in UHV-like settings to suggest that the dominant product may shift from
a mixture of CO and CO$_3^{2-}$ to mainly CO$_3^{2-}$ as the binding
of ionic decomposition products becomes less favorable (e.g., on
thicker oxide coatings).  This prediction may be transferrable to natural
SEI films arising entirely from electrolyte decomposition.
Atomic-scale surface heterogeneity is found to affect EC decomposition,
with Li$^+$ ions at the surface playing a facilitating or ``catalytic'' role.
Our work paves the way for novel future experimental studies in UHV settings.

\section*{Acknowledgement}
 
We thank John Sullivan, Steve Harris, Na Sai, Anatole von Lilienfeld, and
David Rogers for useful discussions, Michael Siegal and Donald Overmyer
for the nanoporous carbon samples, and Xingcheng Xiao for sharing
Ref.~\onlinecite{xcx} prior to publication.  Sandia National Laboratories
is a multiprogram laboratory
managed and operated by Sandia Corporation, a wholly owned subsidiary of
Lockheed Martin Corporation, for the U.S.~Deparment of Energy's National
Nuclear Security Administration under contract DE-AC04-94AL85000.  
KL (apart from the work on lithium metal modeling)
was supported by Nanostructures for Electrical Energy Storage (NEES), an Energy
Frontier Research Center funded by the U.S. Department of Energy, Office of
Science, Office of Basic Energy Sciences under Award Number DESC0001160.

\section*{Supporting Information Available}
Further information are available regarding reorganization energies of
EC in bulk liquid, vertical excitation energies of EC at LiAlO$_2$/liquid EC
interfaces, DFT/PBE predictions of adiabatic bond-breaking barriers, and
discussions of possible LiAlO$_2$ stoichiometry on ALD Al$_2$O$_3$ coatings
upon cycling power.  This information is available free of charge via the
Internet at {\tt http://pubs.acs.org/}.

\newpage
 
\newpage
 
\begin{references}
\bibitem{deloc_ex}
An oft-cited example of
self-interaction error manifests itself in the dissociation of H$_2^+$ in
the gas phase.  The lone electron should reside on one of the H's
as the two H nuclei become widely separated, but many DFT functionals
predict half an electron on each.
\bibitem{liquid}
We stress that liquid EC does not exist at T=0~K;
only single adsorbed molecule can be investigated this way.
\bibitem{landau}
For $V_{\rm AB}$=0.23~eV and $\lambda\approx 2~eV$ (predicted for EC
adsorbed on electrode surfaces, Sec.~\ref{nonadiab1}), the ``transmission
coefficient'' $\kappa$=$2P_{\rm LZ}/(1+P_{\rm LZ})$=14.4, where $P_{\rm LZ}$
is the Landau-Zenner factor (see, e.g., Ref.~\onlinecite{dupuis}).  $\kappa>1$
is associated with the adiabatic regime.
\bibitem{note77}
The cDFT ground state for EC$^-$ adsorbed on the thick LiAlO$_2$-coated
electrode is a spin triplet.  To compute $V_{\rm AB}$, which requires
the overlap integral $\langle \Phi_A|\Phi_B\rangle$, a triplet donor
state $|\Phi_A \langle $ is used. If we forcibly move the highest
occupied orbital in the EC$^-$ $|\Phi_B\rangle$ into the minority spin
channel to overlap with the original singlet $|\Phi_A \rangle $, the
resulting $V_{\rm AB}$ is larger by 73\%.  This may be taken as an
estimate of the uncertainty.
\end{references}
\end{document}